\documentclass[]{emulateapj}
\usepackage{apjfonts}
\usepackage{natbib}
\usepackage{graphicx}
\usepackage{epstopdf}
\usepackage{multirow} 
\usepackage{longtable,lscape} 
\usepackage{color}
\usepackage{xspace}
\usepackage{xargs}
\usepackage{url}

\newcommand{\s}{S\'ersic\xspace}
\newcommand{\eg}{e.g.\xspace}
\newcommand{\ets}{early types\xspace}
\newcommand{\lts}{late types\xspace}
\newcommand{\et}{early-type\xspace}
\newcommand{\lt}{late-type\xspace}
\newcommand{\M}{$\log M_\star/M_\odot$}

\shorttitle{Galaxies in transition}
\shortauthors{Vulcani et al.}

\begin{document}
\title{From blue star-forming to red passive: galaxies in transition in  different environments} 

\author{Benedetta Vulcani\altaffilmark{1}}
\author{Bianca M. Poggianti\altaffilmark{2}}
\author{Jacopo Fritz\altaffilmark{3,4}}
\author{Giovanni Fasano\altaffilmark{2}}
\author{Alessia Moretti\altaffilmark{2, 5}}
\author{Rosa Calvi\altaffilmark{5}}
\author{Angela Paccagnella\altaffilmark{5}}

\affil{\altaffilmark{1}Kavli Institute for the Physics and Mathematics of the Universe (WPI), Todai Institutes for Advanced Study, the University of Tokyo, Kashiwa, 277-8582, Japan}
\affil{\altaffilmark{2}INAF - Astronomical Observatory of Padova, 35122 Padova, Italy}
\affil{\altaffilmark{3}Sterrenkundig Observatorium Vakgroep Fysica en Sterrenkunde Universiteit Gent, Krijgslaan 281, S9 9000 Gent, Belgium}
\affil{\altaffilmark{4}Centro de Radioastronom\'ia y Astrof\'isica, CRyA, UNAM, Campus Morelia, A.P. 3-72, C.P. 58089 Michoac\'an, Mexico}
\affil{\altaffilmark{5}Dipartimento di Astronomia, vicolo Osservatorio 2, 35122 Padova, Italy}

\begin{abstract}
Exploiting a mass complete ($M_\ast$$>$$10^{10.25}M_\odot$) sample at 0.03$<z<$0.11 drawn from the 
Padova Millennium Galaxy Group Catalog (PM2GC), we use 
the $(U-B)_{rf}$ color and morphologies to characterize galaxies, in particular those that show signs of an ongoing or recent transformation 
of their star formation activity and/or morphology - green galaxies, red passive  late types, and blue star-forming early types.
Color fractions depend on mass and only for $M_\ast$$<$$10^{10.7}M_\odot$ on environment. 
The incidence of red galaxies increases with increasing mass, and, for $M_\ast$$<$$10^{10.7}M_\odot$, decreases 
toward the group outskirts and in binary and single galaxies. The relative abundance of green and blue galaxies is 
independent of environment, and increases monotonically with galaxy mass. We also inspect  galaxy structural parameters, star-formation 
properties, histories and ages and propose an evolutionary scenario for the different subpopulations.
Color transformations are due to a reduction and suppression of SFR in both bulges and disks which does not noticeably affect galaxy structure. Morphological transitions are linked to an enhanced bulge-to-disk ratio due to the removal of the disk, not to an increase of the bulge. Our modeling suggests that green colors might be due to star formation histories declining with long timescales, as an alternative 
scenario to the classical ``quenching'' processes.
Our results suggest that galaxy transformations
in star formation activity and morphology depend neither on environment nor on being a satellite or the most massive galaxy of a halo. 
The only environmental dependence we find is the higher fast quenching 
efficiency in groups giving origin to post-starburst signatures.
\end{abstract}



\section{Introduction}
Galaxy color and structure are key observables in extragalactic astronomy for understanding the formation and evolution of galaxies, and they are the consequence of all physical processes at work.

The local population of galaxies consists roughly of two types, and their frequency correlates with the environment: red galaxies,  which on the whole are characterized by larger stellar masses, bulge-dominated morphologies, 
are predominant in dense regions, while blue galaxies, with 
a disk-dominated morphology, are preferentially found in low density regions \citep{blanton03, kauffmann03, kauffmann04, baldry04, balogh04, brinchmann04}. Since only relatively small amounts of ongoing star formation make a galaxy appear blue, the color bimodality basically reflects star formation quenching: in general, red galaxies have had their star formation quenched, while blue galaxies are still forming stars. However, a non-negligible fraction of red galaxies are clearly edge-on disc objects that owe their color to an enhanced extinction, and a small fraction of blue galaxies might have already stopped their activity (see,  \eg, \citealt{bamford09, schawinski09}). 

The bimodality can originate both from a priori differences set beforehand, the so-called nature scenario, or from environmentally driven processes taking place during the evolution of  galaxies, the so-called nurture scenario.  As discussed in \cite{delucia12}, trying to separate the two frameworks and differentiate their role in driving galaxy evolution  might be  an ill posed task, since 
 they are strongly and physically connected. 
According to the $\Lambda$CDM model, as time goes by, smaller structures merge to form progressively larger ones. This hierarchical growth implies that the fraction of galaxies located in groups progressively increases since $z$$\sim$1.5, and at $z$$\sim$0 most galaxies are found in groups \citep{huchra82, eke04, berlind06, knobel09}. It is therefore important to understand the 
 role of the group environment in boosting galaxy transformations from blue to red colors and from late- to \et morphologies.
Color and morphological fractions are very different functions of environment at low-z \citep{bamford09}. Being both sensitive to stellar mass, at fixed stellar mass, color is also highly sensitive to environment, while morphology displays much weaker environmental trends (see also \citealt{kauffmann04, blanton05, christlein05, weinmann09, kovac10}).  

The existence of a variety of ``sub-populations'' of galaxies whose color does not correspond to what is expected based on their morphology (red late-types, blue early-types, etc) suggests that galaxy transformations from blue to red must occur on significantly shorter time-scales than  transformations from late to \et.

However, what drives the observed trends remains still not fully understood.
Numerous processes may be responsible for the dependence of galaxy properties on environment (\citealt{boselli06} and references therein).
The extreme local densities reached within cluster cores enable efficient ram pressure stripping of the galaxy cold gas on timescales of a few Myr \citep{gunn72, abadi99}. On the other hand, 
galaxy-group interactions like ``strangulation'' 
can remove warm and hot gas from a galaxy halo, efficiently cutting off the star formation gas supply \citep{larson80, cole00, balogh00, kawata08}. 
Halos can play a role through tidal forces and dynamical friction. Galaxy-galaxy harassment at the typical velocity dispersion of bound groups and clusters may also result in star-formation quenching \citep{moore96}. 
Shock heating in massive halos can prevent accretion of cold gas that would feed star formation. Interactions and mergers can also apply torques that drive gas inward, perhaps feeding and then exhausting star formation or a central black hole.

Radial trends of galaxy properties (\eg, colors, morphologies) as a function of distance from the halo center can be observable effects of these processes responsible for  the galaxy transformations. They have been extensively studied in galaxy clusters, where \eg, a strong radial dependence in the star-formation rate is observed \citep{hashimoto99, balogh99, lewis02, balogh04, tanaka04,vdlinden10}.

Another way to gain insight into the  physical processes is to study those galaxies whose morphological type places them on one side of the bimodality but whose star formation identifies them with the other.
Blue early-type galaxies with high current star formation rates (0.5$<$SFR$<$50 $M_\odot/yr$) or a recently stopped star formation activity are one example (\eg, \citealt{kannappan09, ferreras09, schawinski09}). These galaxies tend to live in lower density environments than red sequence \ets and make up $\sim$6\% of the low-$z$ general field early-type galaxy population.
They might be early-type galaxies previously on the red sequence that are undergoing an episode of star formation due to the sudden availability of cold gas (``rejuvenated''), making them leave the red sequence before rejoining it. The gas might become available after a merger. If merging occurs between gas-rich galaxies, it may produce a larger amount of star formation (wet mergers), and transform disc galaxies into elliptical galaxies \citep{lin08}.

Another example of objects in transition is given by red \lts. Their distribution displays a clear trend with both local density and group-centric distance: their fraction increases with increasing local density or decreasing group-centric distance,
but at very high densities or in the cores of groups the red \lt fraction declines sharply \citep{bamford09, masters10, vanderwel09}. 
Many of them have some ongoing star formation, and are reddened by dust extinction (\eg \citealt{gallazzi09, wolf09}). 
They might be the result of 
 quite gentle processes (e.g., galaxy-galaxy interactions, interaction with the inter galactic medium (IGM),  harassment, strangulation, 
bar instabilities) that might 
allow the existence of the spiral structure even shutting down the star formation (e.g. \citealt{walker96, skibbasheth09, skibba09}). 

Aim of this study is characterize in detail the incidence of galaxies of different types in the local universe, and depict objects in transition, whose analysis will help us to shed light on the processes acting on galaxies and the time scale needed to galaxies to transform from one type to the other. 

First, we study  colors and  morphologies of galaxies in different environments. Since many galaxy characteristics are interrelated \citep{cowie96, gavazzi96, blanton03, kauffmann03, brinchmann04, baldry04}, we study the correlations for each property independently while fixing other variables. When constraining environmental effects, we perform the analysis in different galaxy stellar mass bins. Galaxy colors and morphologies also correlate, so we split galaxies simultaneously by color and morphological types to distinguish between processes that affect star formation rates and structural properties differently.
Second, we focus on objects in transition  
and study in detail their properties, 
with the aim to understand the evolutive scenario of these galaxies. 

The analysis has been carried out using a cosmology with ($\Omega_m$, $\Omega_\Lambda$, $h$) = (0.3, 0.7, 0.7), Vega magnitudes (unless otherwise stated) and a \cite{kr01} Initial Mass Function (IMF).

\section{The data set and data sample}\label{data}
We use the Padova-Millennium Galaxy and Group Catalogue (PM2GC - \citealt{rosa}), 
consisting of a spectroscopically complete sample of galaxies
at $0.03\leq z \leq 0.11$ brighter than $M_B=-18.7$.
This sample is
sourced from the Millennium Galaxy Catalogue (MGC; \citealt{liske03,driver05}), a B-band contiguous equatorial
survey of $\sim$$38\, {\rm deg}^2$ complemented by a 96\% spectroscopically
complete survey down to B = 20 and it 
is representative of the general field population
in the
local Universe.

By applying a friends-of-friends (FoF) algorithm, \cite{rosa}  identified 176 galaxy groups with at least three members with $M_B<$-18.7 in the redshift range $0.04$$\leq$$z$$\leq$0.1.
A galaxy is considered a group member if its spectroscopic redshift lies within $\pm 3\sigma$ (velocity dispersion) from the median
group redshift and if it is located within a projected distance of 1.5$R_{200}$ from the group geometrical center, where $R_{200}$\footnote{The $R_{200}$  values are computed 
from the velocity dispersions as in \cite{finn05}.} 
is defined as the radius delimiting a 
sphere with interior mean density 200 times the critical 
density of the universe at that redshift,
and is commonly used as an approximation of the group
virial radius.
Galaxies that have no neighbors or just one with a projected mutual distance 0.5 $h^{-1}$ Mpc and a redshift within 1500 $km \,s^{-1}$ are considered ``single''  or ``binary-system'' galaxies, respectively.

Applying  a FoF to the \cite{delucia07} semi-analytic model \citep{vulcani14}, we found that 80\% of our group/binary systems/single galaxies span a halo mass range of $10^{12}-10^{14} M_\odot$/$10^{11.4}-10^{12.6} M_\odot$/$10^{11.2}-10^{12.3} M_\odot$. 

Rest-frame absolute magnitudes are computed using INTERREST \citep{taylor09} from the observed SDSS photometry. 
The code uses a number of template spectra to carry out the interpolation from the observed photometry in bracketing bands (see \citealt{rudnick03}). Rest-frame colors are derived from the interpolated rest-frame apparent magnitudes. 

Stellar masses are estimated following the \cite{bj01} relation \citep{rosa}, which correlates the stellar mass-to-light ratio with the optical colors of the integrated stellar population, using the B-band photometry taken from the MGC, and the rest-frame $B - V$ color computed from the Sloan $g - r$ color corrected for Galactic extinction:
\begin{equation}\label{bj}
\log_{10}(M/L_{B})=-0.51+1.45(B-V)
\end{equation}
valid for a Bruzual \& Charlot model with solar metallicity and a \cite{salpeter55} IMF 
(0.1-125 $M_{\odot}$). Then,  they are converted  to a \cite{kr01} IMF, adding -0.19 dex to the logarithmic value of the masses \citep{cimatti08}.
The typical uncertainty on  mass estimates is 0.2-0.3 dex (for details and comparisons with external  estimates  refer to \citealt{rosa, bia_sd}).
The  sample is complete for  \M$>$10.25, corresponding to the mass of the faintest and reddest galaxy ($M_B$ =-18.7, $B-V$=0.9) at our redshift upper limit ($z$=0.1), as described in \cite{rosa_morph}. 

Star formation rates (SFR) and histories (SFH) are derived by fitting the spectra with the spectrophotometric model fully described in \cite{fritz07, fritz11}. The MGC spectroscopic database of PM2GC galaxies consists of SDSS, 2dFGRS and MGCz spectra \citep{driver05}, the latter taken with the same instrument and setup of the 2dFGRS. Choosing always the highest quality spectrum available, SDSS spectra are preferred when possible (86\% of the sample), alternatively 2dFGRS spectra when available (12\%) and MGCz spectra in the remaining cases. 
In the \cite{fritz07} model, all the main spectrophotometric features (i.e. the continuum flux and shape, the equivalent widths of emission and absorption lines) are reproduced by summing the theoretical spectra of simple stellar populations (SSP) of 12 different ages (from $3 \times 10^6 $ to $\sim 14 \times 10^9$ years).  
The spectral analysis allows to derive an estimate of the SFRs at different cosmic times and of the average age of the stars in a galaxy. 

Morphologies are determined using MORPHOT \citep{fasano12}, an automatic tool designed to reproduce as closely as possible the visual classifications. MORPHOT adds to the classical CAS (concentration/asymmetry/clumpiness) parameters a set of additional indicators derived from digital imaging of galaxies and has been proved to give an uncertainty only slightly larger than the eyeball estimates. It was applied to the B-band MGC images to identify ellipticals, lenticulars (S0s), and later-type galaxies \citep{rosa_morph}.

Structural parameters are taken from  the public catalog of the MGC database \citep{allen06}, and were derived from 2D surface brightness profile fitting of MGC-BRIGHT galaxies using GIM2D \citep{simard02}. We use the catalog obtained adopting a bulge-disk decomposition model using a \s bulge plus exponential disc.  As  described in \cite{allen06}, the components were required to have a common spatial centre, but their luminosities were independent, allowing the calculation of a bulge-to-total (B/T) luminosity ratio for each galaxy.
 We refer to the original paper for additional details. 

In this paper, we consider two galaxy classes and treat separately the most massive galaxy (MMG) of each structure and all the other galaxies, called satellites, regardless of environment. With this definition, all single galaxies are treated as MMGs, while binary systems are split into one MMG and one satellite. 
In addition, we remove AGNs, since their presence  might alter results.  To do that, we match our sample with the latest AGN catalog from SDSS,\footnote{{\url https://www.sdss3.org/dr10/spectro/spectro\_access.php}} finding that overall 48  galaxies that enter our entire mass complete sample are indeed hosting an AGN. Applying the fraction of AGN obtained ($<4\%$) to the PM2GC subsample without spectra from SDSS, we expect $\sim9$ AGN not detected above our mass completeness limit. We assume that this number is negligible.

The final mass complete sample of galaxies in groups consists of 417 satellites and 165 MMGs. 
The final binary system and single galaxy samples consist of 228 (170 of which are MMGs) 
and 418 galaxies respectively.

\section{Galaxy sub-populations}\label{populations}

\begin{figure}
\centering
\includegraphics[scale=0.4]{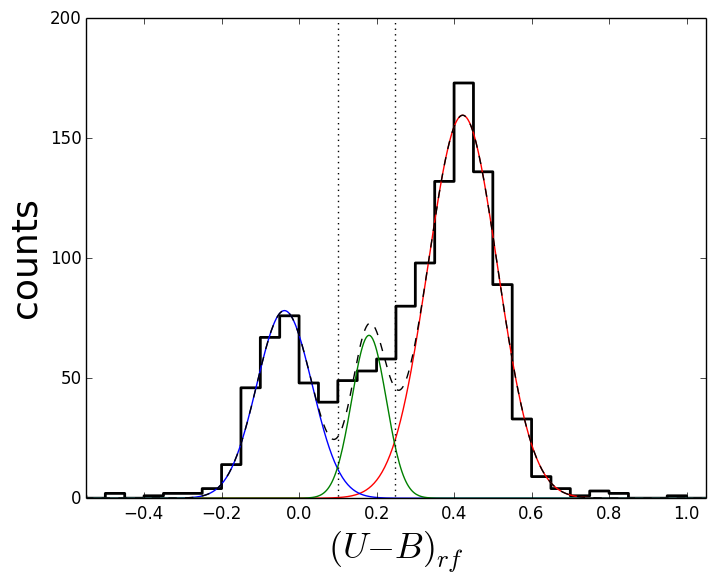}
\caption{Rest frame $(U-B)_{rf}$ color distribution of all galaxies in our mass complete sample. Also plotted are Gaussian fits to the red, blue, green  (solid lines) and total (black dashed line) population. The vertical lines indicate where gaussian fits intersect.\label{colorcut}}
\end{figure}
\begin{figure*}
\centering
\includegraphics[scale=0.3]{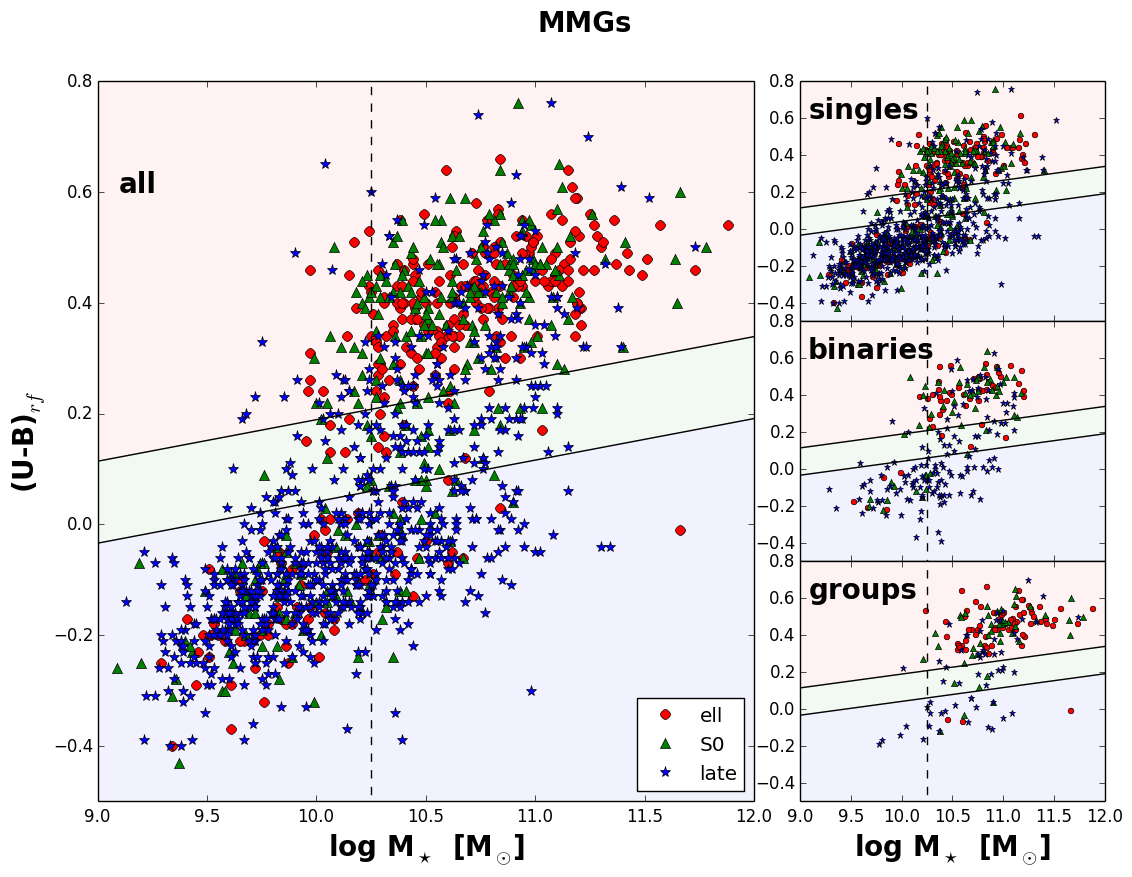}
\includegraphics[scale=0.3]{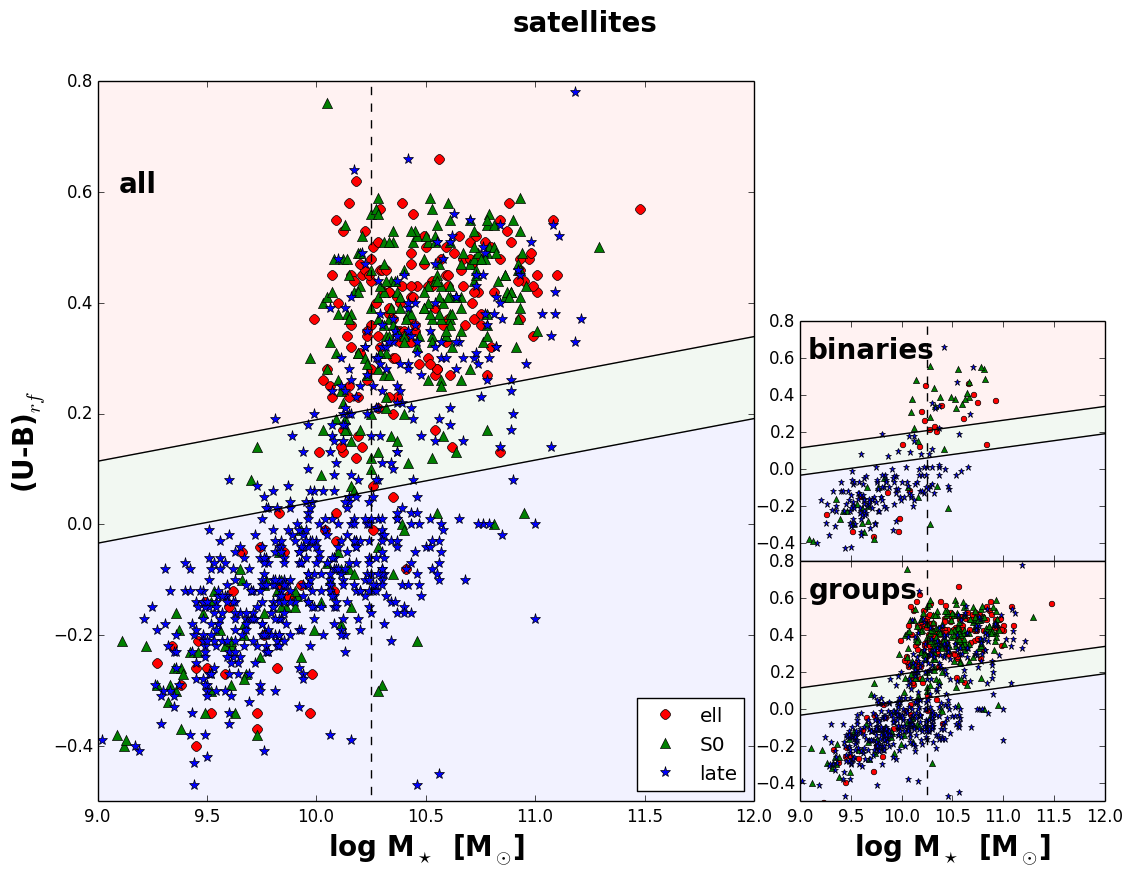}
\includegraphics[scale=0.3]{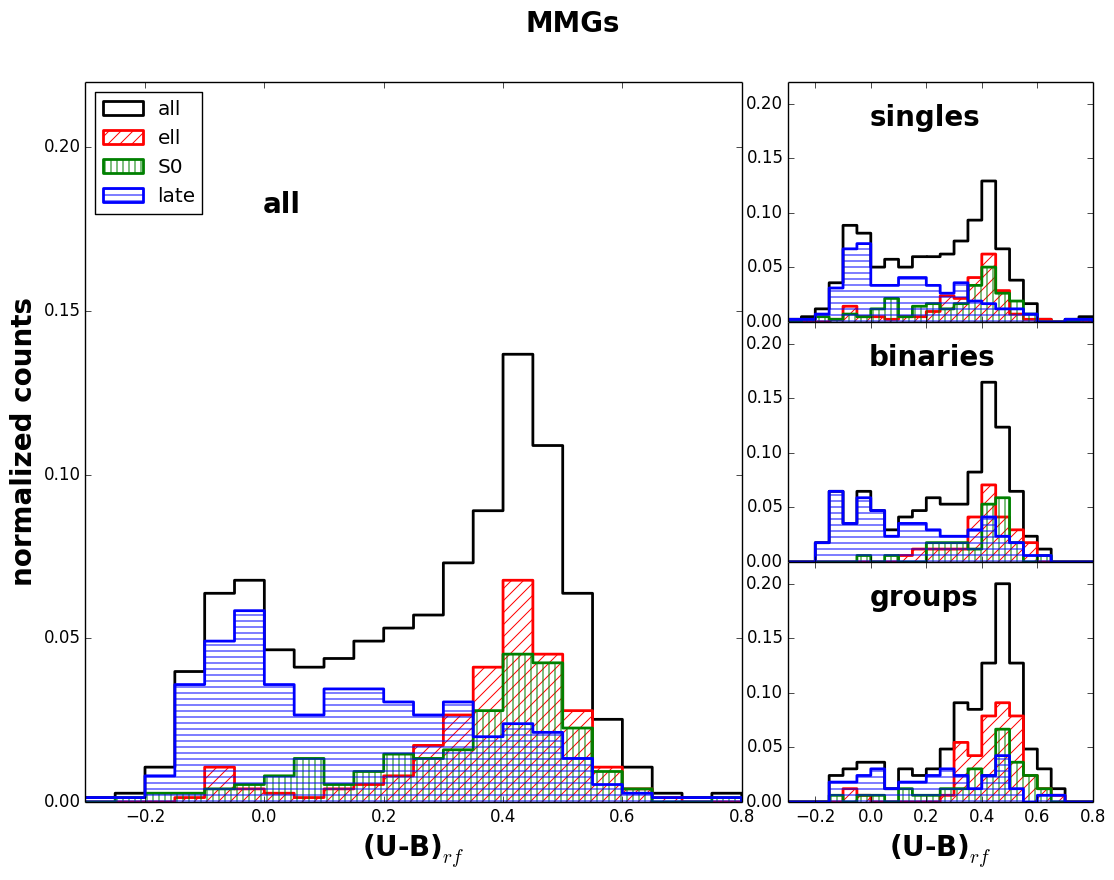}
\includegraphics[scale=0.3]{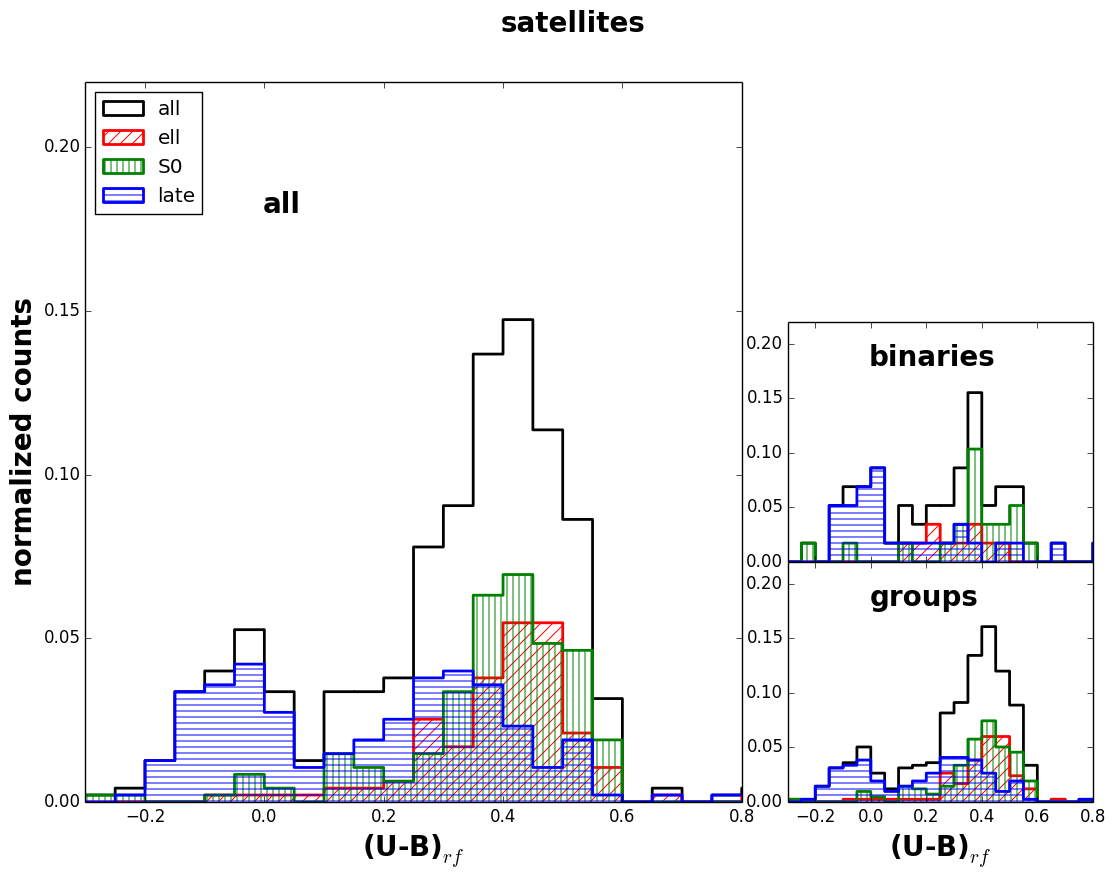}
\caption{Rest frame $(U-B)_{rf}$- mass relation (upper panels) and  normalized rest frame $(U-B)_{rf}$ color distribution (bottom panels) for galaxies of different classes (left panels: MMGs , right panels: satellites) and morphologies in the different environments (shown in the smaller windows). Red circles and histograms: ellipticals; green triangles and histograms: S0s; blue stars and histograms: \lts. In the upper panels, the black dashed vertical line represents the mass completeness limit, the black solid lines show the separation between red, green and blue galaxies (see text for details).  \label{fig_col}}
\end{figure*}

Broadly speaking, based on color, galaxies can be subdivided into two populations: red and blue.
To start, we plot the  $(U-B)_{rf}$ distribution of all galaxies above the mass completeness limit. Figure \ref{colorcut} highlights the presence of the well known bimodality, but also of an emerging third population, located in between the main peaks. We will refer to these as ``green'' galaxies. 
We fit the entire distribution with three Gaussian functions, to define the boundaries of the green population so it has the minimal overlap with the red and blue sequences. The position of the peaks depends on mass 
and redshift, so in principle, we should fit the color distribution at different stellar masses and in different redshift bins and then interpolate to find a relation that depends on stellar mass, color, redshift. Since no significant evolution is expected between $z$$\sim$0.1 and $z$$\sim$0.04, we neglect the redshift dependence. Given that our mass range is quite small, we can not compute the dependence of the cut on stellar mass, and we adopt the slope of the color-mass relation presented in \cite{peng10} (0.075). To compute the zero-points of the relations that separate different colors, we focus on the mass bin $M_\star$ =$10^{10.5-11}M_\odot$ and find the values where the blue and green, green and red gaussians intersect each other. Galaxies are assigned to the red sequence if their rest-frame color obeys 
\begin{small}
\begin{eqnarray}\label{col}
(U-B)_{Vega} \geq 0.075\times \log M_\ast-0.561
 \end{eqnarray} \end{small}
 to the blue cloud if their rest-frame color obeys 
\begin{small}
\begin{eqnarray}\label{col1}
(U-B)_{Vega} \leq 0.075\times \log M_\ast-0.709
 \end{eqnarray} \end{small}
and to the green valley if they are in between the two cuts. This is visually shown in the upper panel of Fig. \ref{fig_col}, where
 we plot also galaxies that do not enter our mass complete sample, just to show the general trends. 

Table \ref{tab_frac} presents the 
percentage of galaxies of different types and colors in different environments for the mass complete samples. The majority of galaxies are red, and more so for satellites  ($\sim$70\%) than MMGs ($\sim$60\%). Green galaxies are 14\% of all MMGs and 9\% of all satellites, while blue galaxies are 26\% of MMGs and 19\% of satellites. 
\begin{table*}
\caption{Percentage of galaxies of different types above the stellar mass completeness limit in different environments. \label{tab_frac}}
\centering
\setlength{\tabcolsep}{3pt}
\begin{tabular}{ll|ccc ccc|ccc ccc}
\hline
\hline
 & & \multicolumn{6}{c|}{{\bf MOST MASSIVE GALAXIES}} & \multicolumn{6}{c}{{\bf SATELLITES}} \\
\hline
\hline
 &&\multicolumn{12}{c}{{\bf ALL GALAXIES}} \\
& & \multicolumn{3}{c}{{\bf COLOR}} 	& 	\multicolumn{3}{c|}{{\bf MORPHOLOGY}} &\multicolumn{2}{c}{{\bf COLOR}} & 	\multicolumn{3}{c}{{\bf MORPHOLOGY}}  \\ 
 & & $red$ & $green$ &$blue$ 			& $ell$ & $S0$ & $late$ 				&  $red$ & $green$ &$blue$ 		& $ell$ & $S0$ & $late$ \\ 
\hline
\parbox[t]{2mm}{\multirow{5}{*}{\rotatebox[origin=c]{90}{{\bf COLOR}}}}  & $all$ &  60$\pm$3 &14$\pm$2	 & 26$\pm$2	 & 28$\pm$2 	 & 25$\pm$2  & 47$\pm$3  &  72$\pm$3 &9$\pm$2  	  &19$\pm$3  & 25$\pm$3  & 35$\pm$3 & 40$\pm$3 \\ 
 																		& $red$ & 			&		  	& 			& 25$\pm$2 		& 19$\pm$2 	& 16$\pm$2  & 			&		  	& 			& 23$\pm$3 	& 30$\pm$3 & 19$\pm$3 \\ 
 																		& $green$& 			&			&			& 1.2$\pm$0.6 	& 3$\pm$1 	& 9$\pm$2  &			&		 	&			& 1$\pm$1 	& 3$\pm$1  & 5$\pm$1\\
 																		& $blue$ & 			&			&	  		& 2$\pm$1  		& 3$\pm$1   & 21$\pm$2	&  		   	&   		& 			& 1$\pm$1   & 2$\pm$1  & 16$\pm$2\\ 
\hline
\hline
 &&\multicolumn{12}{c}{{\bf SINGLE GALAXIES}} \\
& & \multicolumn{3}{c}{{\bf COLOR}} 	& 	\multicolumn{3}{c|}{{\bf MORPHOLOGY}} &\multicolumn{2}{c}{{\bf COLOR}} & 	\multicolumn{3}{c}{{\bf MORPHOLOGY}}  \\ 
 & & $red$ & $green$ &$blue$ 			& $ell$ & $S0$ & $late$ 				&  $red$ & $green$ &$blue$ 		& $ell$ & $S0$ & $late$ \\ 
\hline
\parbox[t]{2mm}{\multirow{5}{*}{\rotatebox[origin=c]{90}{{\bf COLOR}}}}  & $all$ &  53$\pm$3 &16$\pm$3	 & 31$\pm$2	 & 24$\pm$3 	 & 26$\pm$3  & 50$\pm$3  &  -- &--  	  &--  & --  & -- & -- \\ 
 																		& $red$ & 			&		  	& 			& 20$\pm$3 		& 17$\pm$3 	& 16$\pm$2  & 			&		  	& 			& -- 	& -- & -- \\ 
 																		& $green$& 			&			&			& 1$\pm$1	 	& 4$\pm$1 	& 10$\pm$2  &			&		 	&			& -- 	& --  & --\\
 																		& $blue$ & 			&			&	  		& 3$\pm$1  		& 4$\pm$1   & 24$\pm$3	&  		   	&   		& 			& --   & --  & --\\ 
\hline
\hline
 &&\multicolumn{12}{c}{{\bf BINARY SYSTEMS}} \\
& & \multicolumn{3}{c}{{\bf COLOR}} 	& 	\multicolumn{3}{c|}{{\bf MORPHOLOGY}} &\multicolumn{2}{c}{{\bf COLOR}} & 	\multicolumn{3}{c}{{\bf MORPHOLOGY}}  \\ 
 & & $red$ & $green$ &$blue$ 			& $ell$ & $S0$ & $late$ 				&  $red$ & $green$ &$blue$ 		& $ell$ & $S0$ & $late$ \\ 
\hline
\parbox[t]{2mm}{\multirow{5}{*}{\rotatebox[origin=c]{90}{{\bf COLOR}}}}  & $all$ &  59$\pm$5 &14$\pm$4	 & 27$\pm$5	 & 26$\pm$5 	 & 21$\pm$5  & 53$\pm$6  &  57$\pm$9 &12$\pm$6  	  &31$\pm$9  & 17$\pm$8  & 36$\pm$9 & 47$\pm$9 \\ 
 																		& $red$ & 			&		  	& 			& 23$\pm$5 		& 19$\pm$5 	& 17$\pm$4  & 			&		  	& 			& 12$\pm$7 	& 29$\pm$9 & 15$\pm$7 \\ 
 																		& $green$& 			&			&			& 2$\pm$2	 	& 0.5$\pm$0.1 	& 11$\pm$4  &			&		 	&			& 5$\pm$5 	& 2$\pm$2  & 5$\pm$5\\
 																		& $blue$ & 			&			&	  		& 0.6$\pm$0.1  		& 1$\pm$1   & 25$\pm$5	&  		   	&   		& 			& 0$^{+1}_{-0}$    & 5$\pm$5  & 26$\pm$9\\ 
\hline
\hline
 &&\multicolumn{12}{c}{{\bf GROUPS}} \\
& & \multicolumn{3}{c}{{\bf COLOR}} 	& 	\multicolumn{3}{c|}{{\bf MORPHOLOGY}} &\multicolumn{2}{c}{{\bf COLOR}} & 	\multicolumn{3}{c}{{\bf MORPHOLOGY}}  \\ 
 & & $red$ & $green$ &$blue$ 			& $ell$ & $S0$ & $late$ 				&  $red$ & $green$ &$blue$ 		& $ell$ & $S0$ & $late$ \\ 
\hline
\parbox[t]{2mm}{\multirow{5}{*}{\rotatebox[origin=c]{90}{{\bf COLOR}}}}  & $all$ &  79$\pm$5 &7$\pm$3	 & 14$\pm$4	 & 42$\pm$6 	 & 26$\pm$5  & 32$\pm$5  &  74$\pm$3 &9$\pm$2  	  &17$\pm$3  & 26$\pm$3  & 34$\pm$3 & 40$\pm$3 \\ 
 																		& $red$ & 			&		  	& 			& 40$\pm$6 		& 22$\pm$5 	& 16$\pm$4  & 			&		  	& 			& 24$\pm$3 	& 30$\pm$3 & 20$\pm$3 \\ 
 																		& $green$& 			&			&			& 0$^{+1}_{-0}$ 	& 2$\pm$1 	& 5$\pm$2  &			&		 	&			& 1$\pm$1 	& 3$\pm$1  & 5$\pm$2\\
 																		& $blue$ & 			&			&	  		& 2$\pm$2  		& 2$\pm$2   & 10$\pm$4	&  		   	&   		& 			& 1$\pm$1   & 2$\pm$1  & 15$\pm$2\\ 
\hline
\hline
\end{tabular}
\end{table*}

Color fractions strongly depend on environment (singles, binaries, groups), 
and not on galaxy class, that of MMGs and satellites being always compatible within the errors.
Interestingly, green galaxies are always about half as numerous as blue galaxies, showing that their relative number  
is independent both of environment and of the distinction satellites-MMGs (see \S4.1 for more details).

Considering morphologies, \lts dominate both galaxy classes, being $\sim$40\% and $\sim$50\% of all satellites and MMGs, respectively. For both classes, this fraction
decreases going from less to more massive environments.
Ellipticals are instead found in the same proportion among MMGs and satellites. In addition, ellipticals and S0s are found in similar proportions among MMGs, while among satellites S0s are more frequent. These latter trends remain true when individual environments are inspected, except that ellipticals are more numerous than S0s in group MMGs.
Thus, at odds with the color fractions, morphological fractions vary both between groups, binaries and singles {\sl and} from MMGs to satellites in a given environment.

The bottom panels of Fig.~\ref{fig_col} show that, even though a color bimodality exists for both MMGs and satellites, their color distributions are remarkably different: they both show a prominent red peak at $(U-B)_{rf}\sim$0.45 and a secondary blue peak at $(U-B)_{rf}\sim$-0.05. Nonetheless, blue and green galaxies 
are more conspicuous among MMGs than among satellites, suggesting that only the blue fractions depend on environment and class.
Going from the least toward the most massive environments, the two peaks of the distributions shift toward redder colors, the peak at blue colors gets less prominent
while the number of galaxies with red color increases, indicating again that in groups galaxies are most likely red. This is valid for both MMGs and satellites. 

Splitting galaxies by morphology, 
among MMGs, \lts show an unimodal distribution with a long tail toward green and red colors, while among satellites they present a bimodal distribution, with a second peak almost as important as the first one, centered at quite red colors, and only a few green galaxies. 
Late types are the galaxies that show the most noticeable variation with the galaxy class and environment. 
Ellipticals and S0s are mostly red, but green and blue ellipticals and S0s 
are a non-negligible fraction.
In MMGs, ellipticals and S0s show similar color distributions peaked around $(U-B)_{rf}\sim$0.4. 
Both distributions show a tail of galaxies with blue colors.
In satellites, instead, S0s are the most important population forming the red peak.

This analysis highlights the variation of the incidence of each sub-population with environment, and the well known fact that there is no one-to-one correspondence between color and morphology, i.e. not all \lts have blue colors, and not all ellipticals and S0s are red.  

In the following we refer to red \lts, blue early-types and green galaxies of all types as candidate ``transition objects'', because they are likely to have experienced, or being in the process of experiencing a transformation in color and/or morphology,
from star-forming to passive, or viceversa.

Until now we have considered only galaxy colors and morphologies, but these do not
univocally distinguish between star-forming and passive galaxies.
The red \lt sample might be contaminated by the presence of star-forming galaxies highly extincted by dust\footnote{We remind the reader that the \cite{fritz07} model includes a treatment of the dust, hence in principle our derived SSFRs already take into account the possible presence of dust and are corrected for that. In practice, models are not able to detect highly extincted star formation in the optical without the aid of infrared data. The levels of SFR in our sample, however, are not expected to be extreme in the vast majority of cases thus our dust treatment and SFRs can be considered overall reliable.} 
and, viceversa, not all blue or green \ets are necessarily forming stars.
To isolate truly star-forming and truly passive galaxies, we inspect the level of SSFR (=SFR/$M_\star$).\footnote{The typical relative error on the SFFR is 35-40\%.} From now on, we only consider red \lt objects with SSFR$<10^{-12}  yr^{-1}$ (RP late-type) and  blue \et objects with SSFR$>10^{-12} yr^{-1}$ (BSF early-types).

\section{Trends with mass and environment}
\subsection{Mass trends in  different environments}\label{masstrends}
\begin{figure*}
\centering
\includegraphics[scale=0.35]{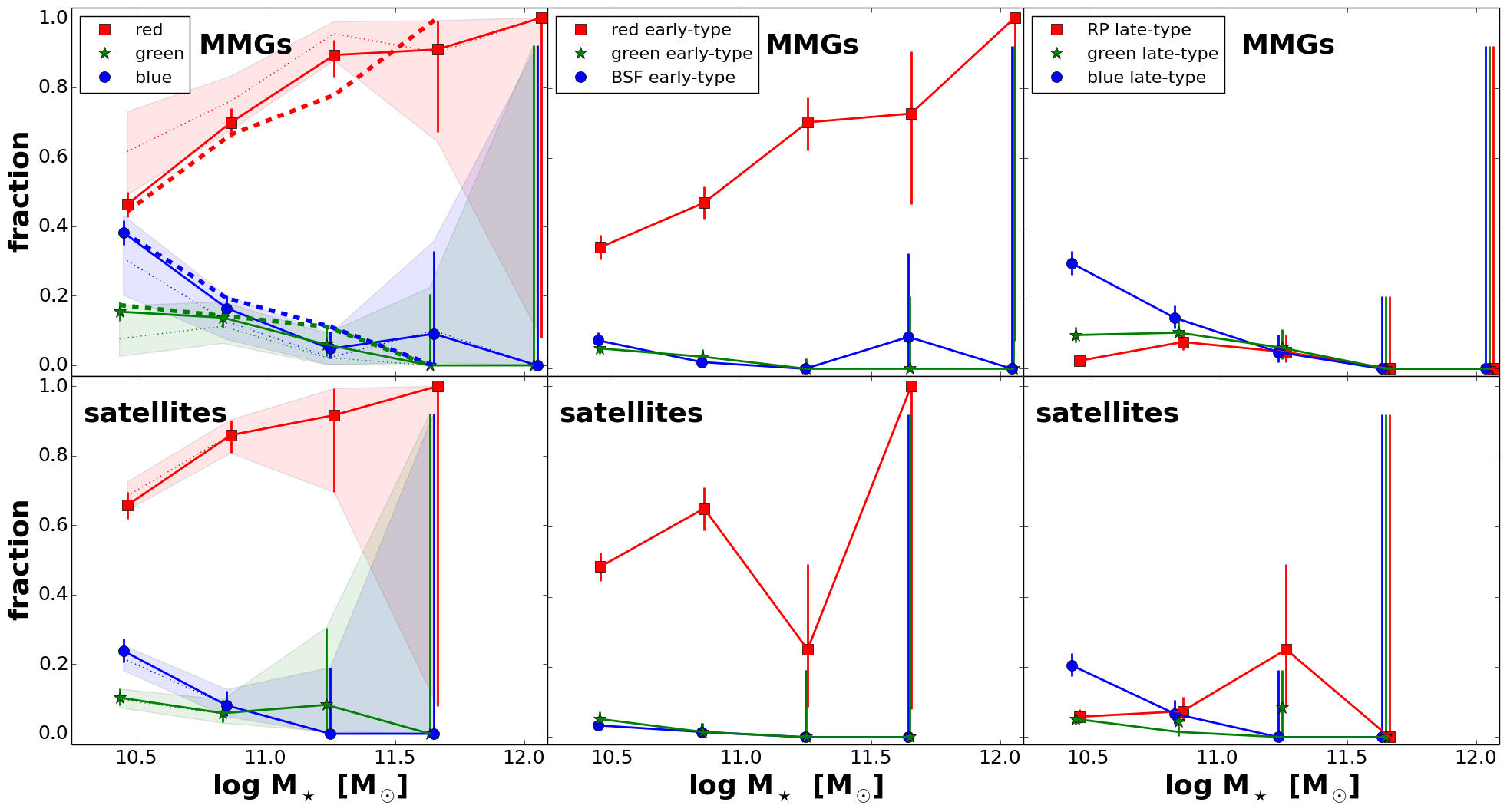}
\caption{Fraction of galaxies as a function of stellar mass for MMGs (upper panels) and satellites (bottom panels). Left panels: color cut; red squares and solid lines: red galaxies, green stars and solid lines: green galaxies, blue circles and solid lines: blue galaxies. Dotted lines and shaded areas represent trends in groups, bold long dashes represent trends in single galaxies. Central panels: color+SSFR+morphological cut for \et galaxies, as indicated in the labels.
Right panels: color+SSFR+morphological cut for \lt galaxies, as indicated in the labels.  Errors are defined as binomial errors \citep{gehrels86}. Small horizontal shifts are applied to make the  plot clearer.
\label{col_frac_mass_dist}}
\end{figure*}

\cite{rosa_morph, rosa_mf} have already shown that in the PM2GC galaxy morphology is linked with both stellar mass and environment. The mass distribution of each morphological type depends on the environment, and in each environment the mass function is different for ellipticals, S0s and \lts. They found that there is little dependence of the morphological fractions on galaxy mass in the range $10.25<M<11.1$, while, at higher masses, this dependence is strong. 

\begin{figure}
\centering
\includegraphics[scale=0.45]{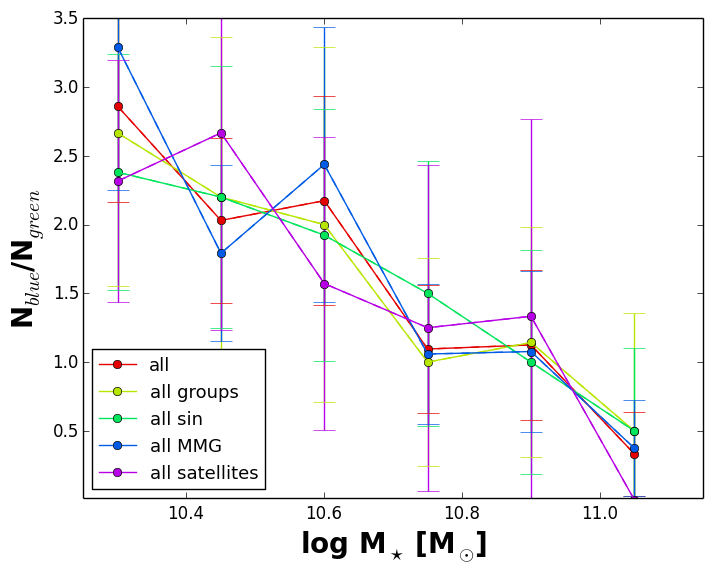}
\caption{Ratio of the number of blue galaxies to the number of green galaxies as a function of stellar mass, for galaxies in different environments and of different class. 
\label{bg_frac}}
\end{figure}

Here we take a complementary approach 
carefully inspecting also  changes in color. 
At first, we apply only the cut in color, 
for MMGs and satellites separately (Fig.\ref{col_frac_mass_dist}).  
Fractions strongly depend both on stellar mass and class. 
As shown in the left panels, the contribution of blue galaxies to the total population steadily decreases with  mass. However, at any  mass, blue galaxies are more numerous among MMGs than among satellites, being 40\% of the entire MMG population at \M=10.4, and $<$25\% in satellites at the same mass. 
Also the fraction of green galaxies decreases with increasing mass, for both MMGs and satellites, though more gently than blue galaxies do. 
In contrast, the incidence of red galaxies increases with mass, and they dominate both galaxy classes at all masses.

Environmental variations of color fractions can also be observed comparing group and single MMGs:
red/blue and green galaxies are less/more frequent among singles than in groups at a given  mass.
No significant differences are detected between group satellites and all satellites, as expected given the fact that 90\% of our satellites are in groups.
On the other hand, we note that considering only groups, color fractions as a function of mass are very similar for MMGs and satellites, indicating that differences between all MMGs and all satellites of a given mass are driven by the single MMGs.

More interesting than the simple green fraction is the number of green galaxies relative to the number of blue galaxies, which probably are evolutionary linked.
Figure \ref{bg_frac} presents the ratio of blue to green galaxies as a function of mass for all galaxies, for single and group galaxies and for MMGs and satellites. Trends are very similar in all cases, being the blue-to-green ratio always a factor 1:2 at high masses, 1:1 at intermediate masses, and 2.5-3:1 at lower masses. We will discuss the meaning of this ratio later, when we will discuss in detail the evolutionary links between green and blue galaxies.

We now apply a joint subdivision in color, morphology and SSFR (central and right panels in Fig.\ref{col_frac_mass_dist}). Here and in the following figures, we consider together elliptical and S0 galaxies (\ets), in order to increase the statistics and improve the readability of the panels.\footnote{We note that for \M$>$11 ellipticals are twice as numerous as S0s (see also \citealt{rosa_morph}).} 
While the fraction of red early-types increases with mass,\footnote{The observed dip in satellite red \ets is probably due to galaxies with a consistent bulge but that have been classified as \lts and small number statistics. Indeed, in Appendix \ref{sersic}, where trends are inspected considering the \s index instead of morphology, the fraction of red bulges steadily increases with mass. } those of BSF and green early-types tend to decrease and are very close to zero at high masses. Also the trends for blue, green and RP late types are different,
steadily declining the former, consistent with being flat with mass the others. This is true both for MMGs and for
satellites.
Moreover, RP \lt, BSF and green \et trends with mass do not change with environment (groups vs. singles, plots not shown).

In \S\ref{disc} we will analyze in detail transition galaxies, but we first focus on trends in groups, where galaxy properties might be driven also by the location of the galaxies within the group.

\subsection{Radial trends in groups}\label{rad_trends}
\begin{figure}
\centering
\includegraphics[scale=0.4]{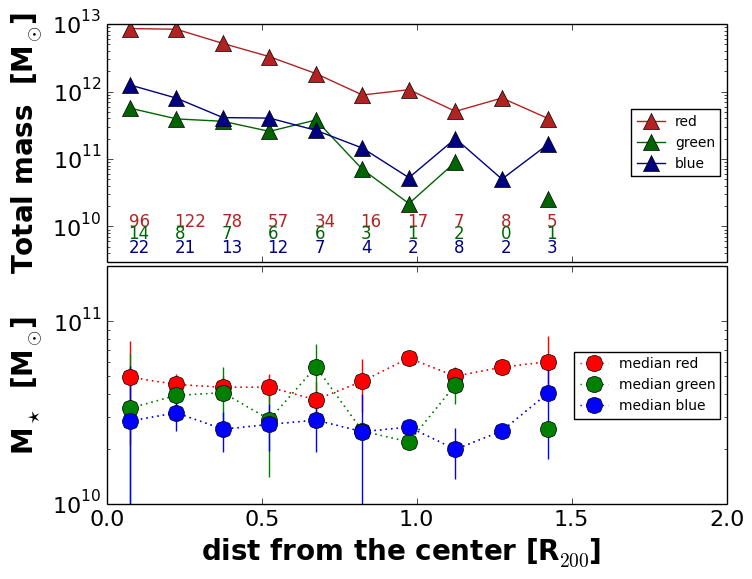}
\caption{Total (upper panel), and median (bottom panel) stellar mass  as a function of group-centric distance for  red, green and blue MMG+satellite galaxies in groups above the mass completeness limit.  Numbers are the number of objects in each bin. \label{dist}}
\end{figure}

\begin{figure*}
\centering
\includegraphics[scale=0.35]{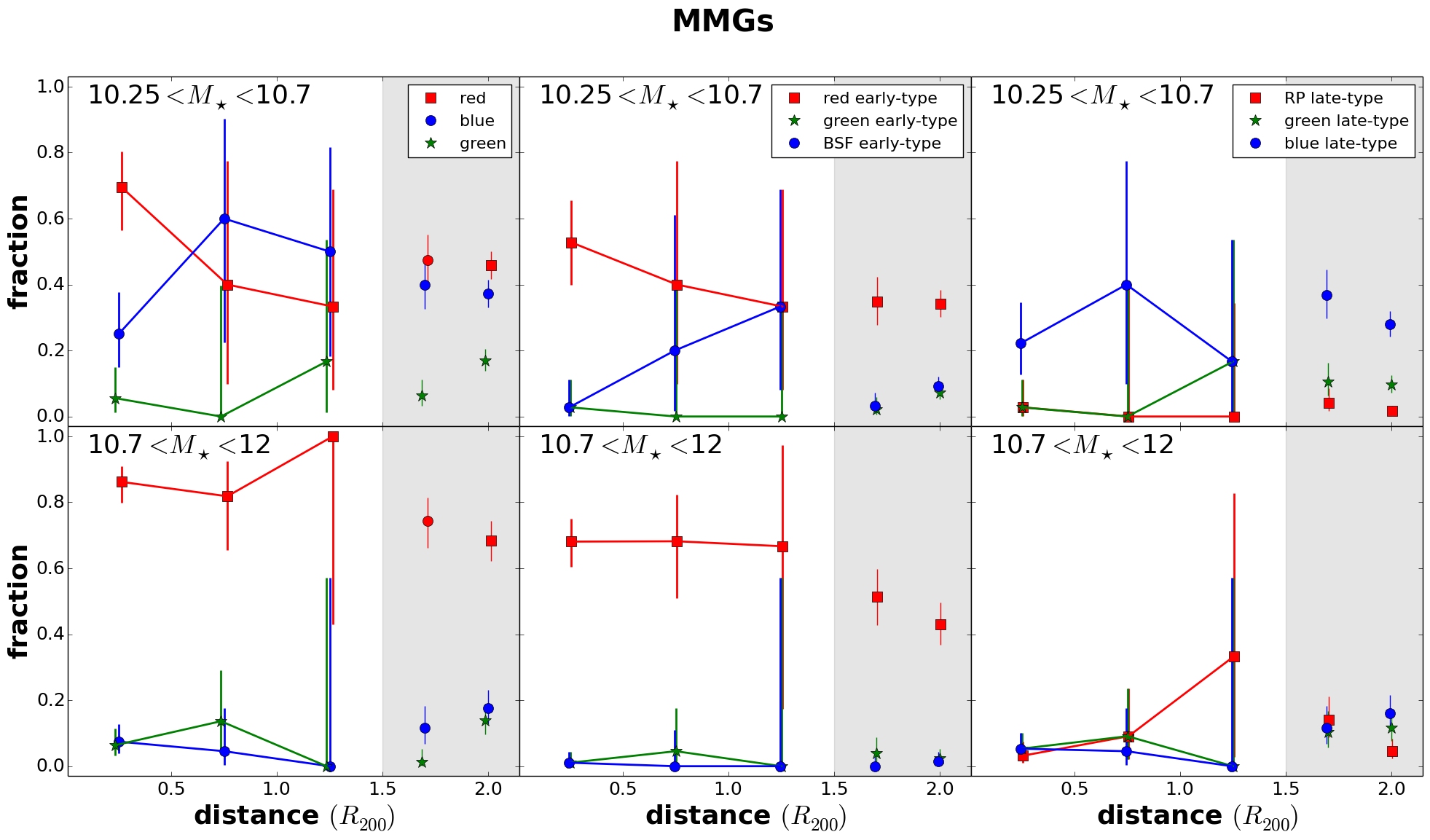}
\includegraphics[scale=0.35]{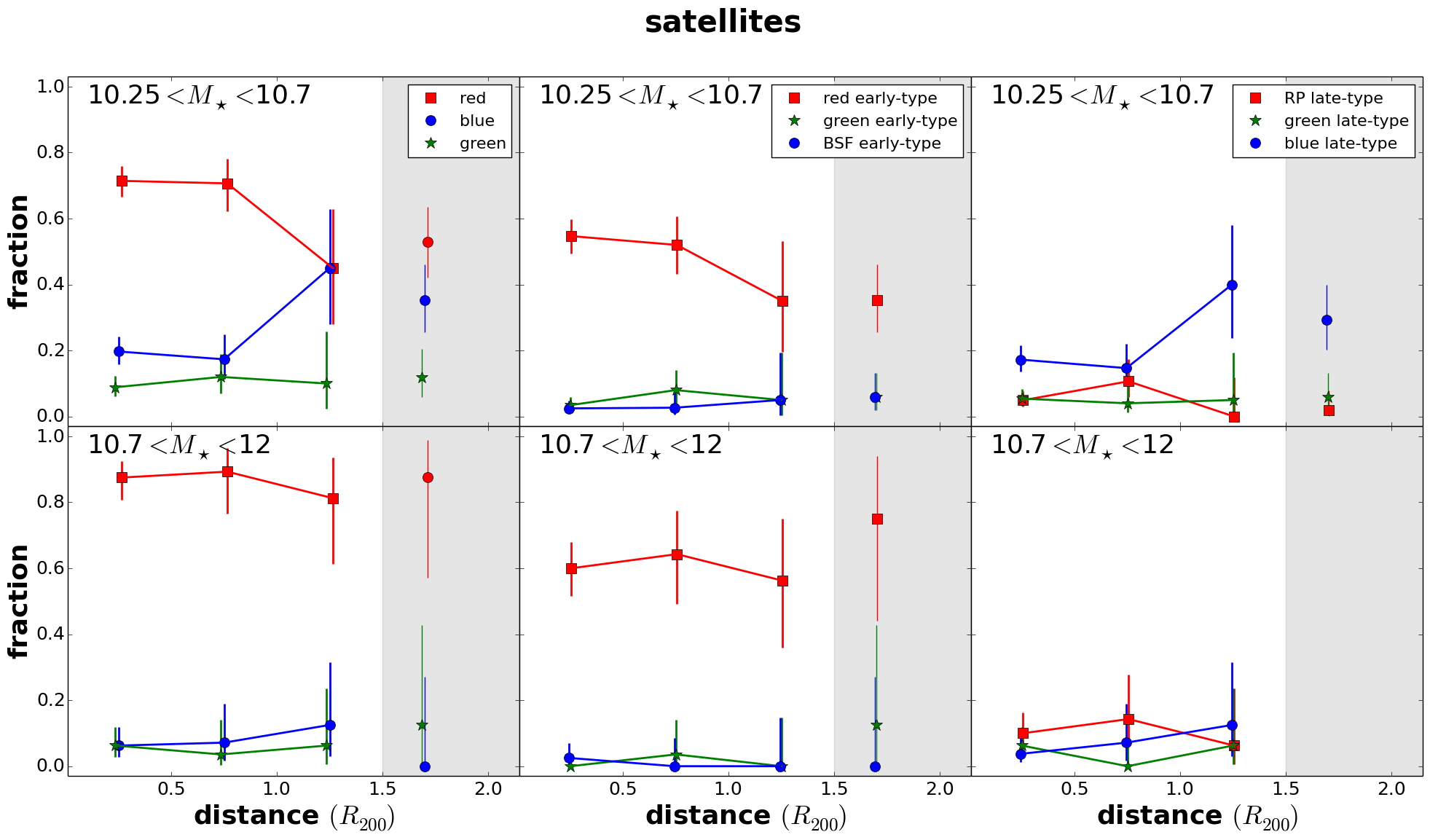}
\caption{Fraction of galaxies as a function of group-centric distance in bins of stellar mass. Upper panels: MMGs, Lower panels: satellites. Left panels: all masses, central panels: \M$<$10.7, right panels: \M$>$10.7.  Upper panels: color cut. Bottom panels: color+morphological cut. Colors and symbols are as in Fig. \ref{col_frac_mass_dist}. Grey area represent galaxies that are not in groups, i.e. binary systems (placed at $r/R_{200}$=1.7) and single galaxies (placed at $r/R_{200}$=2). Small horizontal shifts are applied to make the plot clearer.
\label{col_frac_dist}}
\end{figure*}

In groups there might be an additional parameter that can drive galaxy transformations, that is the position of galaxies within their structure. Indeed galaxies are expected to fall toward halo centers due to dynamical friction, therefore their group-centric distance is related to the time galaxies have spent inside the group \citep{weinmann11, smith12, delucia12}.

We first look for signs of mass segregation with distance. 
Considering MMGs and satellites together, we sum the stellar mass of all galaxies (total mass).  Figure \ref{dist} shows that, both for red, green and blue galaxies, the total mass in galaxies depends on group-centric distance: there is more mass close to the group center than in the outskirts, due to the larger number of galaxies at small group-centric distances. However, the median stellar mass is independent of distance, for galaxies of any color separately, as shown in the bottom panel. Fluctuations in trends of green galaxies are probably due to small number statistics: at $r$$>$0.7$R_{200}$ there are only 7 galaxies. 
While green and blue galaxies present similar average masses, red galaxies are systematically more massive, of a factor of 2.

Therefore, stellar mass and position within a group are not strictly related and might play a different role in driving galaxy transformations. To separate the two contributions, we investigate how fractions depend on group-centric distance in bins of stellar mass. 

We start analyzing galaxy color fractions in group and compare them to binary and single systems.
Figure \ref{col_frac_dist} show qualitatively similar trends for MMGs and satellites, though satellites have better statistics.
At low masses (upper left panel), red galaxies dominate at small group-centric distances, while in the outskirts they are as common as blue galaxies, whose fraction increases with distance. The fraction of green galaxies is consistent within the errors with being constant with distance. At higher masses (lower left  panel) all trends are flat, with red galaxies representing always 
$\sim$75-80\%  of the total population. 
In both mass bins, in binaries,  galaxies of any color are, within the errors, about as common as galaxies of the same color in the group outskirts. The effect of environment is visible only among single galaxies for an excess of blue objects at high masses.

Considering also morphologies and star-forming properties, uncertainties increase, but some trends are still robustly detected. At low masses (upper central and right panels in Fig. \ref{col_frac_dist}), red \ets dominate the populations, but their percentage slightly decreases with distance. In MMGs their decrease is mirrored by BSF \ets, in satellites by blue \lts. The fraction RP \lts,  green \ets and \lts is flat with ditance. At higher masses, all trends are flat, with a possible excess of RP \lts in the outskirts for MMGs. 
We do not detect any significant environmental variation.

\subsection{Possible systematics}
There are several possible biases or other effects in the data to consider in order to ensure the robustness of these results.

First of all, we note that any contamination of the group sample by field galaxies and vice versa, for which we are not applying any correction, will only render the observed trends less prominent. The real, corrected trends therefore would be even more pronounced. 

As described in \cite{rosa}, some of the groups are not fully contained in the narrow strip of the MGC survey and hence suffer from edge problems. 50 of these groups enter our sample, for a total of 313 galaxies. Removing them from our analysis, all the trends are recovered, within the errors (plots not shown).  Since our results are not affected by these groups, we keep them in our sample to have a better statistics.

Our groups show a variety of galaxy richness: the smallest systems host only three members, the biggest one about 60 \citep{rosa}. As a consequence, results might be influenced by the different number of galaxies in groups, so we performed again our analysis splitting galaxies into two bins of richness: those in groups with $N_{gal}<7$ members and those hosted in larger systems.\footnote{The cut is applied considering all members and not only those above the mass completeness limit.} This cut allows us to have almost equally numerous bins.
No signs for a group-richness dependence are evident, except for the fact that richer groups tend to have more massive MMGs. All MMGs in the richest groups are more massive than \M=10.7 (plots not shown).

We also note that the uncertainty associated to the determination of the group centers might be quite large, especially for the groups composed by only 3 objects. Once again, the real, corrected, trends would be more noticeable. 

In our analysis, we have distinguished between MMGs and satellites. However,  MMGs do not always  coincide with the galaxy closest to the group geometrical center.
MMGs are mainly located in the central regions of the haloes (74\% are within $r$=0.5$R_{200}$), but not always at their very center (34\% correspond also to the galaxy closest to the geometrical center).\footnote{In a few cases they are on the edge of the groups, though at least three MMGs at $r$$>$1$R_{200}$ are found in groups that might suffer from edge problems. }  
This can be understood as different stages in the hierarchical clustering process \citep{brough08, pimbblet08}.
 We note that this fraction is in agreement with  
that estimated by \cite{skibba11}, who computed the fraction of brightest non central galaxies as a function of the halo mass, finding it is 25\% in low-mass haloes ($10^{12} h^{-1} M_\odot\leq M \leq 2 \times 10^{13}h^{-1}M_\odot$) and increases with halo mass.
Our choice of contrasting MMGs and all other galaxies 
(common to many studies, \eg, \citealt{weinmann06, weinmann09, skibba07, vdb08, pasquali09, pasquali10}) most likely produces an underestimate of the true differences between centrals and satellites, hence trends might be even more pronounced.

\section{Objects in transition}
In the previous sections, we have investigated the color and morphological fractions as a function of stellar mass and environment. Such analysis has emphasized the presence of objects likely in transition from one type to the other.
Above our mass completeness limit, 4$\pm$1\% of galaxies are BSF \ets (2$\pm$1\% of MMGs, 8$\pm$2\% of satellites), 
4$\pm$1\% are green \ets (4$\pm$1\% of MMGs  and 4$\pm$1\% of  satellites), 8$\pm$1\% are green \lts (9$\pm$1\% of MMGs  and 5$\pm$1\% of  satellites), and 5$\pm$1\% are RP \lts (4$\pm$1\% of MMGs  and 6$\pm$1\% of  satellites).

We now analyze
star formation levels, ages and structural parameters of these galaxies. From now on, we show the results for MMGs and satellites together, to increase the statistics, having checked that no substantial differences exist between the two populations, when the effect of the different mass distribution is considered. 

\begin{figure*}
\centering
\includegraphics[scale=0.4]{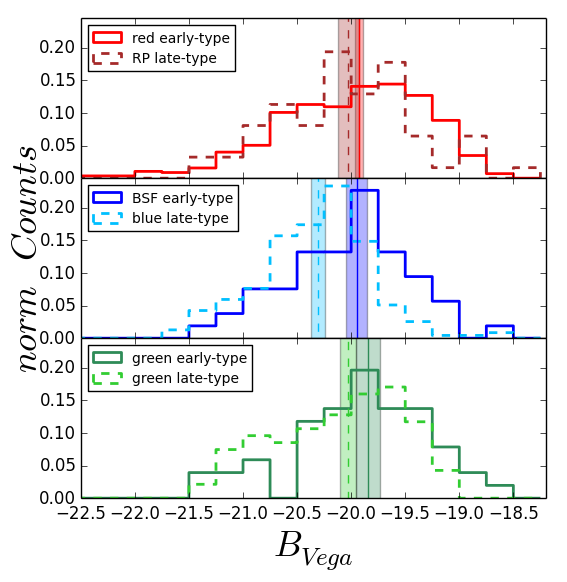}
\includegraphics[scale=0.4]{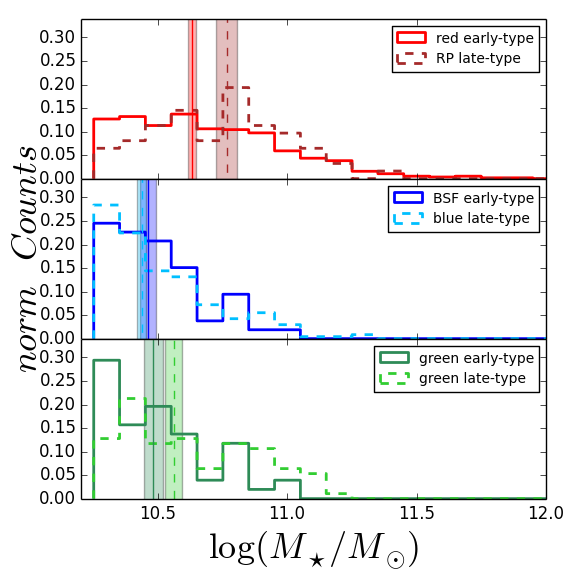}
\includegraphics[scale=0.4]{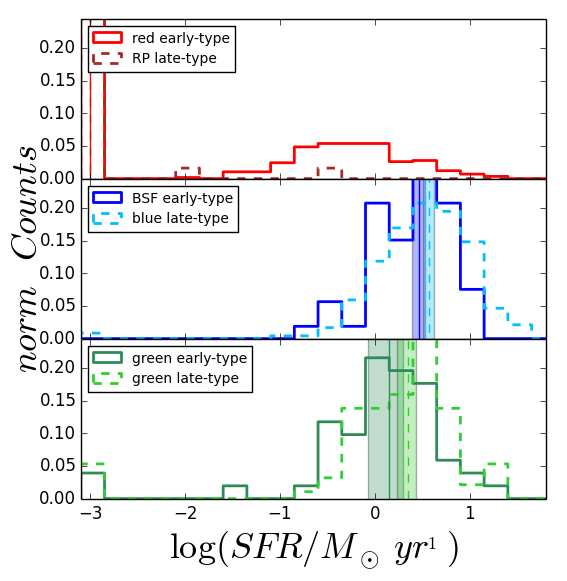}
\includegraphics[scale=0.4]{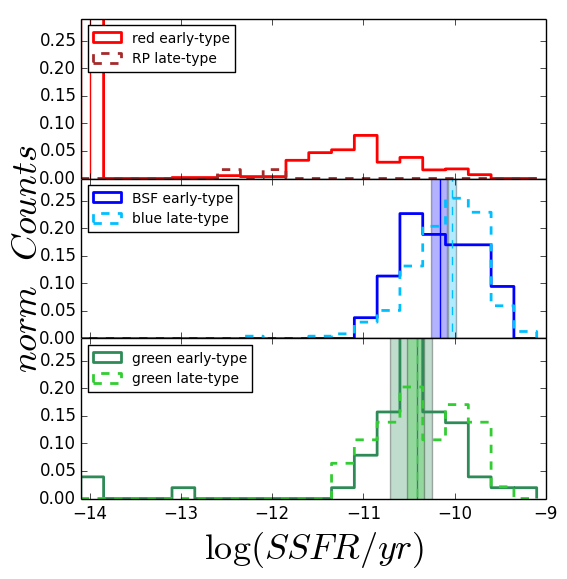}
\includegraphics[scale=0.4]{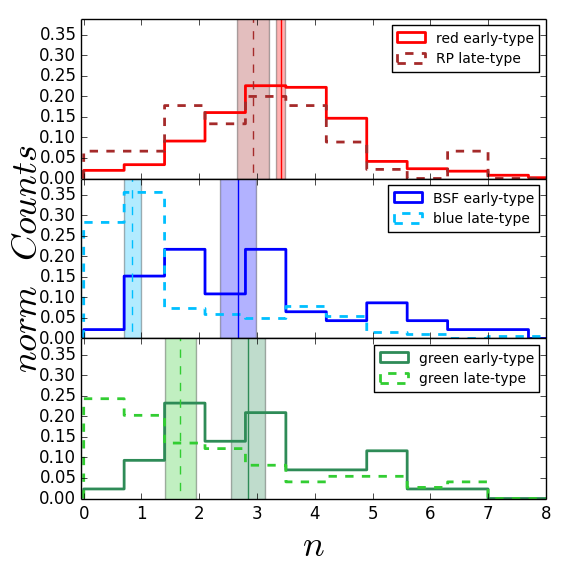}
\includegraphics[scale=0.4]{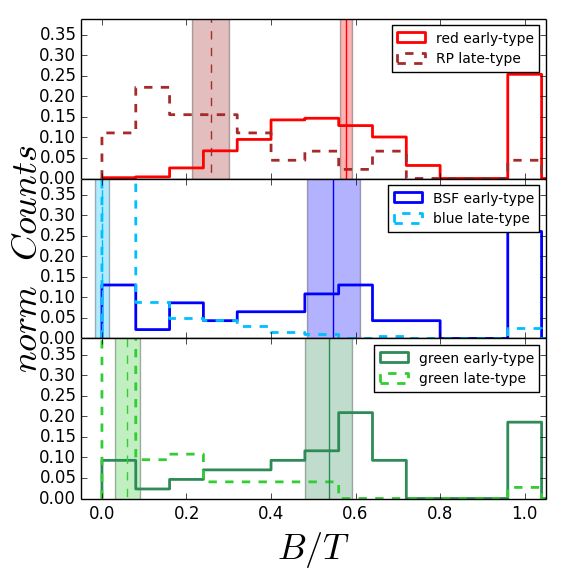}
\includegraphics[scale=0.4]{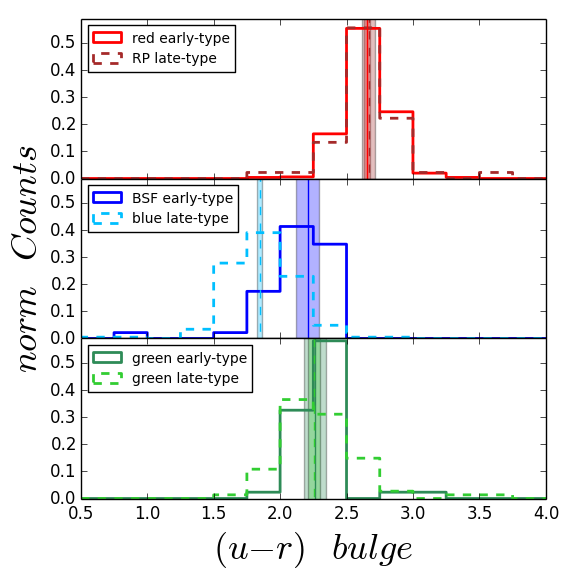}
\includegraphics[scale=0.4]{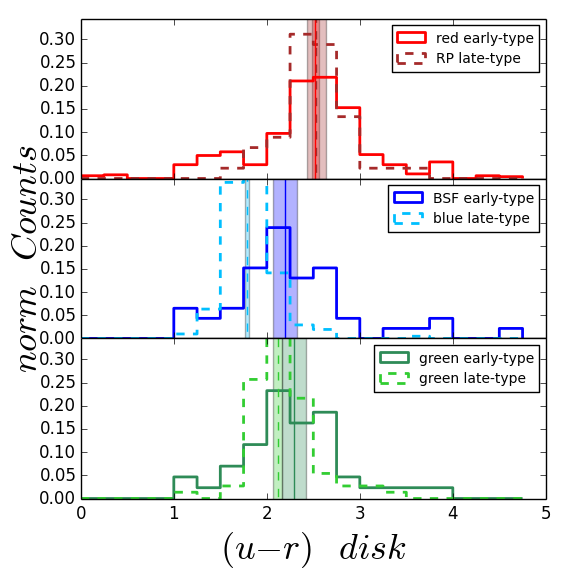}
\includegraphics[scale=0.4]{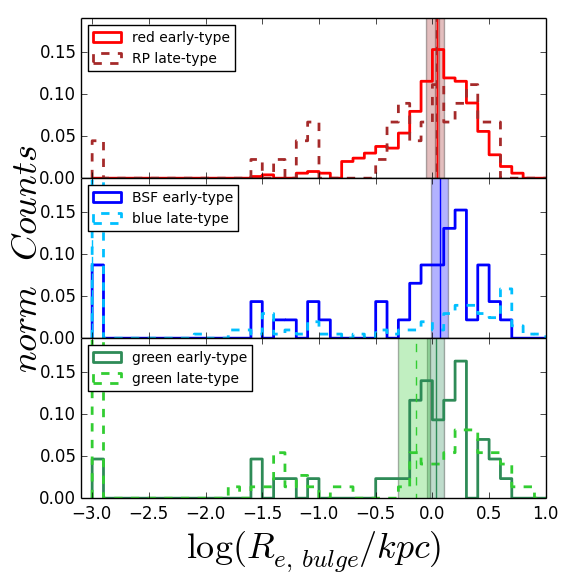}
\includegraphics[scale=0.4]{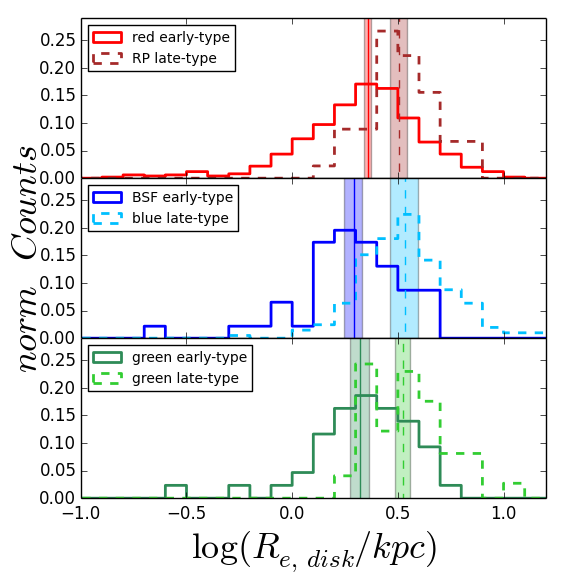}
\caption{Normalized  distributions for BSF \ets (blue), green \ets (dark green), red \ets (red), blue \lts (cyan), green \lts (light green) and RP \lts (dark red), as indicated in the labels. From top to bottom:  B$_{Vega}$ magnitudes, stellar masses, SFRs, SSFRs, 
\s indexes, B/T ratios,  $(u-r)_{rf}$ of bulges,   $(u-r)_{rf}$ of disks, bulge effective radii, disk effective radii. Medians and errors on the medians are also shown. 
\label{et}}
\end{figure*}

\begin{figure*}
\centering
\includegraphics[scale=0.4]{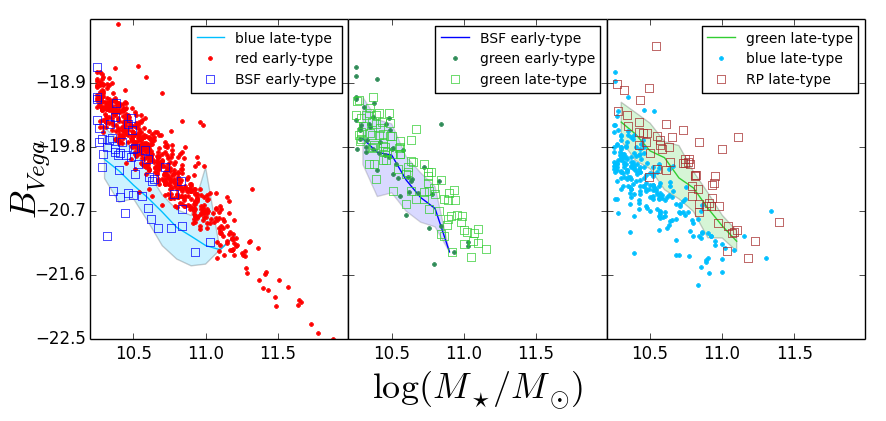}
\includegraphics[scale=0.4]{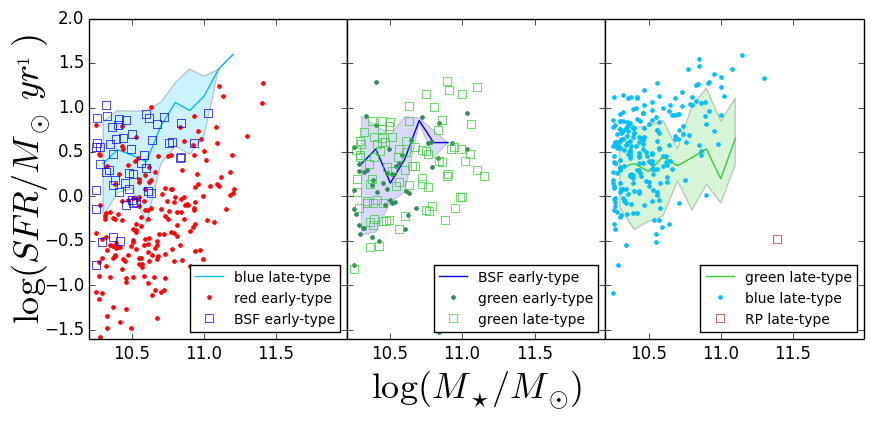}
\includegraphics[scale=0.4]{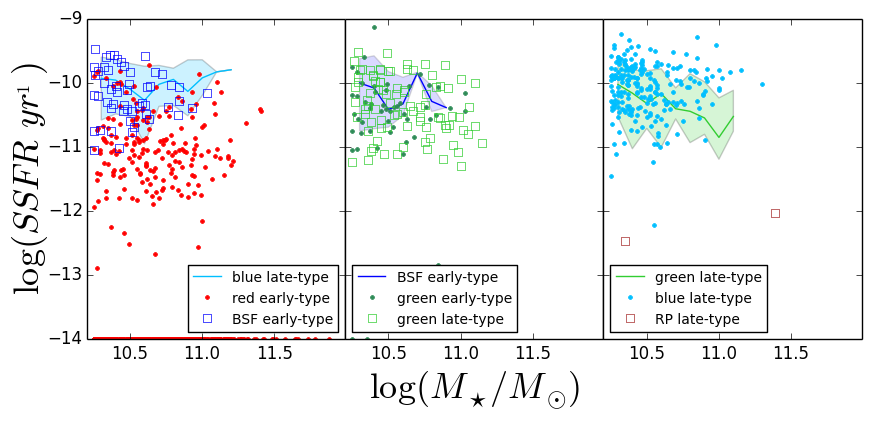}
\includegraphics[scale=0.4]{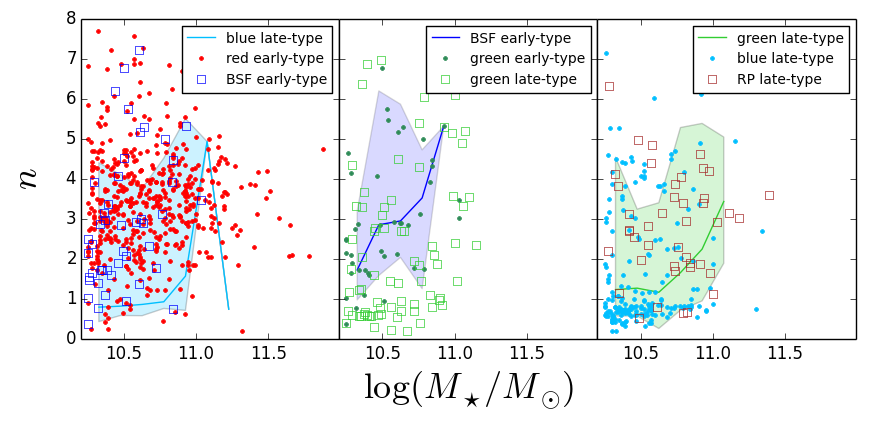}
\includegraphics[scale=0.4]{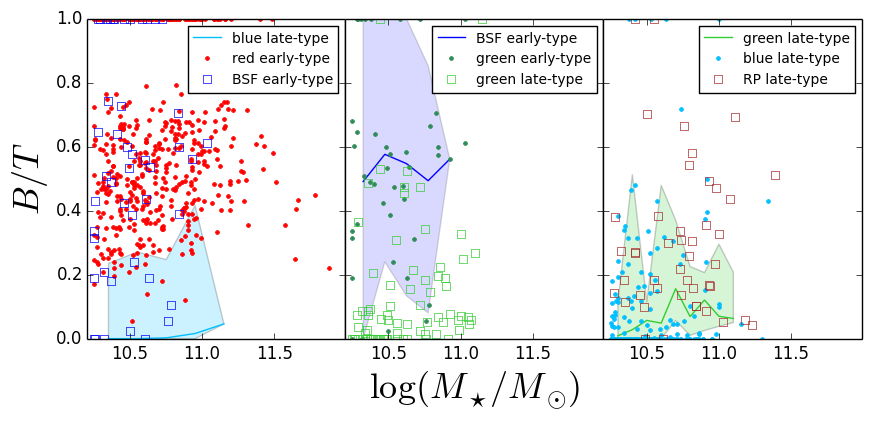}
\includegraphics[scale=0.395]{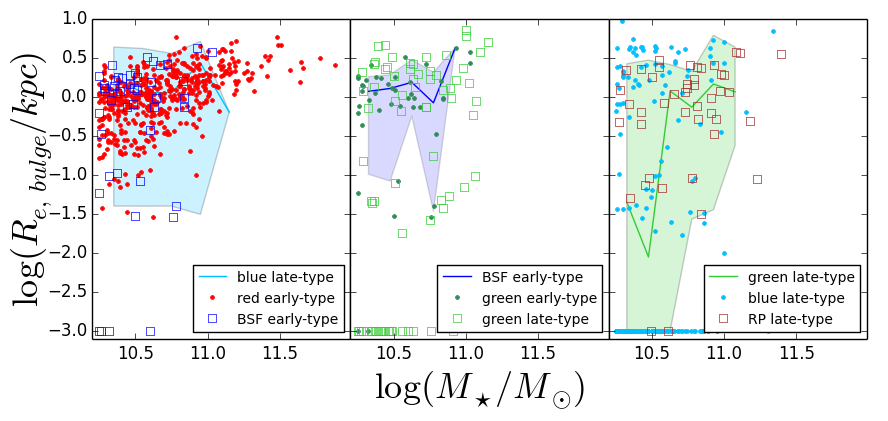}
\includegraphics[scale=0.395]{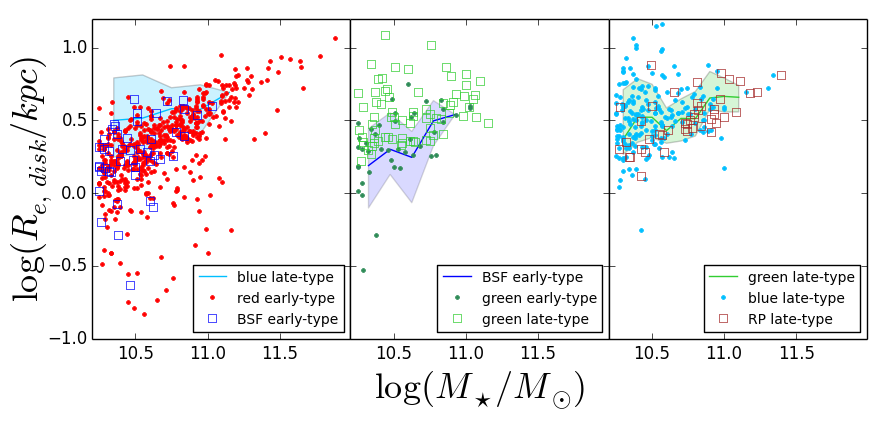}
\includegraphics[scale=0.4]{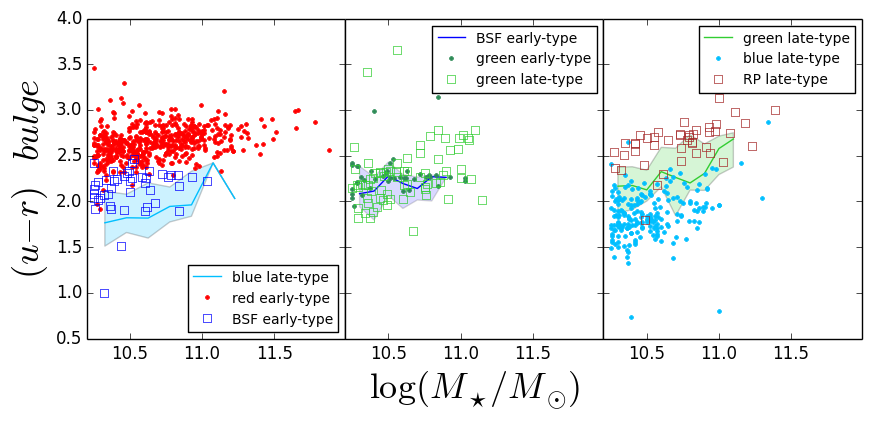}
\includegraphics[scale=0.4]{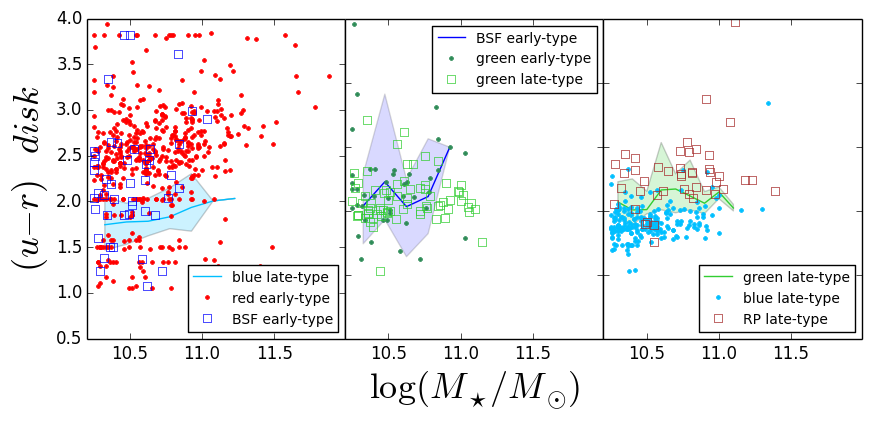}
\caption{Properties as a function of mass for BSF \ets (blue), green \ets (dark green), red \ets (red), blue \lts (cyan), green \lts (light green) and RP \lts (dark red), as indicated in the labels. In each panel, three populations that might be evolutionary linked are shown. For clarity, in each panel one population is represented by its median and a shaded area including its 10-th and 90-th percentiles. From top to bottom:  B$_{Vega}$ magnitudes, SFRs, SSFRs, 
\s indexes, B/T ratios,  $(u-r)_{rf}$ of bulges,   $(u-r)_{rf}$ of disks, bulge effective radii, disk effective radii. 
\label{et_mass}}
\end{figure*}

Figure \ref{et} presents the normalized distributions of the most relevant quantities for the objects in transition, compared to their normal counterparts: $M_{B_{Vega}}$ magnitudes, stellar masses, SFRs, SSFRs, 
\s indexes, B/T ratios,
(u-r) rest-frame color of bulges and disks,  radii of bulges and disks. 
Figure \ref{et_mass} presents the same quantities as a function of stellar mass, in order to show that the different mass distributions of the samples are not fully responsible for the differences observed in the
stellar population and structural parameters. 
\begin{table*}
\caption{Characteristic numbers of objects in transition, compared to normal galaxies of the same morphology. 
\label{tab_trans}}
\centering
\setlength{\tabcolsep}{3pt}
\begin{tabular}{lcccccc}
\hline
\hline
Quantity &\multicolumn{2}{c}{{\bf red}}&\multicolumn{2}{c}{{\bf green}} &\multicolumn{2}{c}{{\bf blue}}\\
		& early-type 	& passive late-type 				& early-type 	& late-type 				& SF early-type 	& late-type \\
\hline
number &576 &62  &51 &94 & 50 & 236\\
$\langle$B$_{Vega}$ $\rangle$ & -19.93$\pm$0.04 & -20.0$\pm$0.1 & -19.84$\pm$0.1& -20.03$\pm$0.07 &-19.9$\pm$0.1 & -20.31$\pm$0.06\\
$\langle$\M$\rangle$ & 10.63$\pm$0.04 & 10.76$\pm$0.05 &10.48$\pm$0.05 & 10.56$\pm$0.05 & 10.46$\pm$0.05 & 10.43$\pm$0.03\\
$\langle$SFR$\rangle$ ($M_\odot yr^{-1}$)& 0$\pm$0 & 0$\pm$0 &1.4$\pm$0.5&2.2$\pm$0.5 & 2.9$\pm$0.5 & 3.6$\pm$0.5\\
$\langle$SSFR$\rangle$$^{\alpha}$ ($yr^{-1}$)& (9$\pm$3)10$^{-12}$ & (6$\pm$3) 10$^{-13}$ &(4$\pm$2) 10$^{-11}$ & (4.2$\pm$0.8)10$^{-11}$&(6$\pm$1)10$^{-11}$ & (9.4$\pm$0.9)10$^{-11}$\\
$\langle$n$\rangle$ & 3.41$\pm$0.08 & 2.9$\pm$0.3 & 2.8$\pm$0.3&1.7$\pm$0.2 & 2.7$\pm$0.3 & 0.8$\pm$0.1\\
$\langle$B/T$\rangle$ &0.58$\pm$0.01& 0.26$\pm$0.04&0.54$\pm$0.06 & 0.06$\pm$0.03 & 0.55$\pm$0.06 &0$\pm$0\\
$\langle R_e \, (bulge)\rangle$ (kpc)& 1.10$\pm$0.05 & 1.07$\pm$0.2& 1.1$\pm$0.2& 0.7$\pm$0.2&1.2$\pm$0.2 & 0$\pm$0\\
$\langle R_e \, (disk)\rangle$ (kpc) &2.28$\pm$0.09 & 3.2$\pm$0.3&2.1$\pm$0.2 & 3.3$\pm$0.3 &2.0$\pm$0.2 &3.4$\pm$0.5\\
$\langle$(u-r) bulge$\rangle$$^\beta$ & 2.65$\pm$0.02& 2.67$\pm$0.05 & 2.26$\pm$0.09 &2.26$\pm$0.05&2.21$\pm$0.09 &1.84$\pm$0.22\\
$\langle$(u-r) disk$\rangle$$^\beta$ &2.52$\pm$0.04&  2.5$\pm$0.1& 2.3$\pm$0.1 & 2.11$\pm$0.05&2.2$\pm$0.1 &1.79$\pm$0.02\\
\hline
\multicolumn{4}{l}{\footnotesize{$^{\alpha}$Values computed only with galaxies with SSFR$\neq$0}}\\
\multicolumn{4}{l}{\footnotesize{$^{\beta}$Colors are in the AB system}}\\
\end{tabular}
\end{table*} 
Table \ref{tab_trans} gives the mean values of the aforementioned quantities, for all the different sub-populations.  The quoted uncertainties on the median are estimated as $1.253 \sigma/\sqrt{N}$, where $\sigma$ is the standard deviation about the median and $N$ is the number of galaxies in the sample under consideration \citep{rider60}.

In the following, we directly compare the properties of the objects in transition to those of the subpopulations that might share a similar history with them.

\subsection{Blue star-forming early types}
We now compare BSF \ets to red \ets and blue \lts (Fig. \ref{et} and left panels of Fig. \ref{et_mass}). Since ellipticals and S0s are expected to be characterized by different properties, in Appendix \ref{S0ell} we report the differences between these two populations. 

The mass distribution of BSF \ets is very similar to blue \lts, while red \ets are systematically more massive. 
At any given mass, BSF \ets are on average brighter than their red counterparts, but fainter than blue \lts. 

The SFR-mass (and SSFR-mass) relations are similar for BSF \ets and blue \lts, but that of BSF \ets is truncated at SFR$\sim$10 $M_\odot$, while that of blue \lts is not. 
In contrast, the SFR-mass relation for the subset of red \ets that are star-forming is well below the standard relation at all masses.

BSF \ets resemble their red counterparts in the $n$-mass plane.
They span the whole range of B/T ratio, from pure disks to pure bulges. Both their $n$ and $B/T$ distributions are very different from those of blue \lts.

The bulge and disk colors of BSF \ets
are in between those of their red counterparts and blue \lts. 
Bulges lie on the upper edge of the corresponding blue \lts color-mass relation, while most of the disks lie just above it. 
In addition, while most of the bulges follow the same mass-size relation in BSF and in red \ets, we detect a subpopulation that is more than an order of magnitude smaller. This bimodality in bulge size is seen also in the BSF \lt population, where many galaxies have no bulge. Disks in BSF galaxies have similar sizes to those in red galaxies and trace the lower edge of the blue \lt relation. 

We also analyzed the concentration of light (as derived in \citealt{abraham94}), mass and luminosity weighted ages\footnote{Following the definition of \cite{cid05}, the luminosity-weighted (mass-weighted) age is computed by weighting the age of each SSP composing the integrated spectrum with its bolometric flux (mass).} and the fractions of mass enclosed in the bulge (plots not shown).
BSF are less concentrated and systematically younger than red \ets. In addition, BSF and red \ets of similar mass have a quite similar fraction of mass in the disk, but a few BSF galaxies host a higher fraction of mass in the disk than their red counterparts.

\subsection{Green \ets}
\begin{figure}
\centering
\includegraphics[scale=0.4]{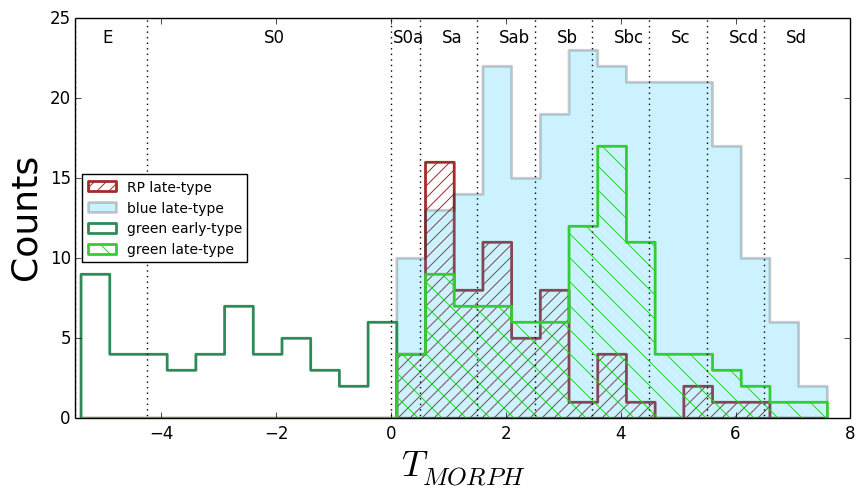}
\caption{Finer morphological distribution of all (MMG+satellites) green, RP \lts and blue \lts,  above the mass completeness limit. 
\label{green_morph}}
\end{figure}

Figure \ref{green_morph} shows the morphological distribution of  green galaxies and blue and RP \lts. Focusing on the first population, it emerges it is mainly composed of late types but includes also some
ellipticals and S0s. In Appendix \ref{S0ell} we will discuss in details the differences between these two populations, here we just focus on all \ets.

The properties of green \ets can be compared to those of green \lts and BSF \ets (Fig. \ref{et} and central panels of Fig. \ref{et_mass}).
Green \ets  and BSF \ets have  similar mass distributions and maximum mass (\M$\sim$11.1), while green \lts are slightly more massive. 
Green \ets and BSF \ets have  similar \s indexes, B/T ratios,
similar color in the bulge and the disk, similar sizes of bulge and disks. At any fixed stellar mass, they only show a light reduction of star formation (SFR and SSFR).

\subsection{Green and red passive \lts}

We now compare green, 
RP 
and blue galaxies with a \lt morphology.
Fig. \ref{green_morph} shows that the RP morphological distribution is clearly skewed to earlier types compared to that of blue \lts, but intermediate and late-type spirals are present too. Green galaxies show an intermediate distribution between the two. 

Also the mass distribution of green galaxies is intermediate between that of blue (less massive on average) and RP (more massive),
even though the maximum mass reached is similar in the three samples (Fig. \ref{et}).

In addition, most green \lts are not completely quenched yet: their SSFR distribution indicates that most of them are still forming stars, only at a lower average rate than blue galaxies of similar mass. Only a few ($\sim$5\%) green \lts are already completely passive. The colors of both their bulges and disks are intermediate between red and blue, indicating that both of these structures are experiencing reduced levels of star formation (Fig. \ref{et} and right panels of Fig. \ref{et_mass}).

As for their structure, green \lts are characterized by somewhat more prominent bulges than blue \lts: on average, they have slightly larger \s indexes,  slightly higher B/T ratios, larger bulges 
than blue galaxies of similar mass.

Focusing on RPs, almost all the distributions  point to differences with blue galaxies: RP late-type galaxies are on average  fainter in the B-band, much more massive,  
have higher $n$ and B/T, 
much redder bulges and disks, quite similar bulge and disk sizes but with the tendency to be on the smaller/more compact side of the size-mass distributions 
than blue late-type galaxies of similar mass. 

In addition, at a given stellar mass,  RP \lts are also noticeably older and contain a higher fraction of mass in the bulge than blue \lts (plots not shown).

\subsection{Post-starburst galaxies}
Another way to understand how blue,  star-forming galaxies turn into red, passive systems is to directly look at their spectra, and seek for signs of such transition.
Adopting the spectral classification defined by \citet[MORPHS collaboration]{dressler99}, we identify $k+a$ galaxies, whose spectra display a combination of signatures typical of both K and A-type stars with strong H$\delta$ in absorption and no emission lines. These features are typical of so called post-starburst/post-starforming galaxies whose star formation was recently (at some point during the last 0.5-1 Gyr) truncated over a short timescale, typically shorter than a few times $10^8$yr.

In our mass limited sample, 4.5\% of all galaxies can be classified as $k+a$. This fraction is independent of environment. However, when considering the relative
number of $k+a$ and blue galaxies, this is much higher in groups ($22\pm6$\%)
than in binaries and singles ($13\pm5$\%), suggesting a higher efficiency
of sudden quenching in groups compared to lower mass haloes.

In our sample, 32 $k+a$ are red (8 
RP \lts), 13 are green and 10 are blue (1 BSF \et).
This shows that the majority of those galaxies that are truncated on a short
timescale cannot be recognized based on color or color+morphology,
but only performing a detailed spectral analysis. The $k+a$ channel is
therefore another transition
channel that needs to be considered to build a complete picture of
galaxy transformations, that affects 4.5\% of all galaxies on a timescale
of $\sim 1$ Gyr.

Finally,
we note that only 6\% green \ets and 10\% of green \lts are $k+a$, confirming
that most of the green galaxies are due to a SF decline on a longer timescale.

\subsection{SFHs of the different populations}
\begin{figure*}
\centering
\includegraphics[scale=0.35]{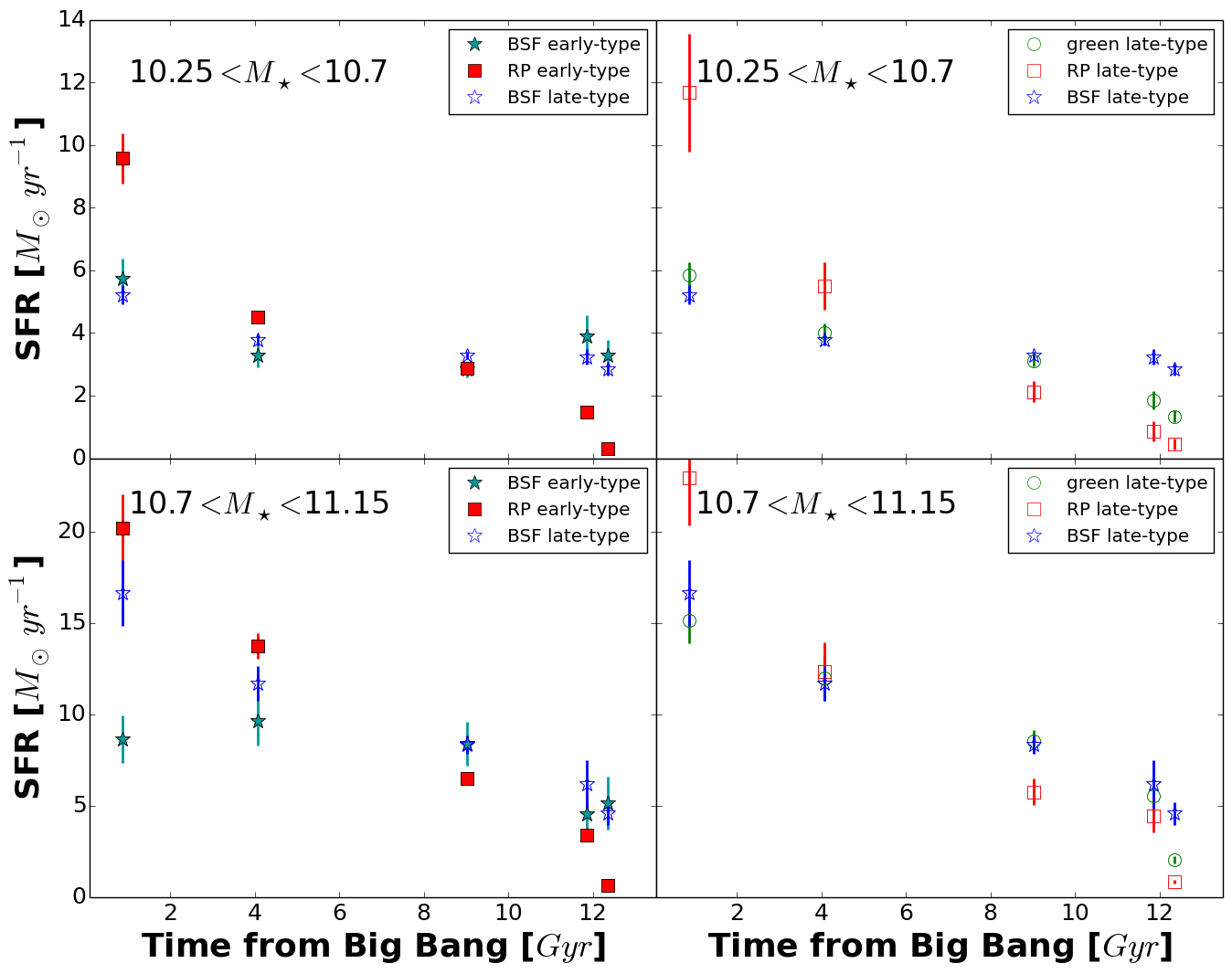}
\caption{Mean SFR as a function of cosmic time for galaxies of different colors and/or star-forming properties, as described in the labels. Top panels: galaxies with \M$<$10.7. Bottom panels: galaxies with 10.7$<$\M$<$11.15. Errors on mean values are obtained using a bootstrap resampling. 
\label{sfh}}
\end{figure*}
We now investigate the SFHs of the different galaxy populations.  
To reduce the effect of the stellar mass in quenching galaxies, we consider two different mass bins: 10.25$<$\M$<$10.7 and 10.7$<$\M$<$11.15. We do not include more massive galaxies due to the lack of statistics. We verified that no significant residual dependence on mass distribution remains in each mass bin.

We compute the mean SFR per galaxy in 5 age intervals  (t=0.877, 4.079 ,9.021, 11.863, 12.345 Gyr) and compare the trends for galaxies of different colors and/or star-forming properties. Errors on mean values are obtained using a bootstrap resampling. Figure \ref{sfh} shows that galaxies with  different properties today are characterized by different average SFHs, which also depend on stellar masses. 

Blue \lts are currently forming stars,\footnote{Blue \lts with $SSFR<10^{-12}\,yr^{-1}$ are 2\% of all blue \lts, and their influence is  negligible.} and the slope of their decline in star formation with time depends on stellar mass, being steeper for more massive galaxies. 
In both mass bins green \lt galaxies have the same SFH of blue \lts at early epochs, the only difference being a significant turn-down of their star formation in the last 1-3 Gyrs (last one or two time bins depending on mass). This is consistent with green \lts being  regular blue \lts
until a few Gyr ago, when they experienced a reduction (not yet a complete halting in most cases) of the star formation activity. Green early- and \lts show very similar SFHs (plot not shown). 

Except for the first bin at high masses, BSF \ets have SFHs identical to blue \lts. 
This is consistent with BSF \ets being blue \lts that suffered an alteration of their morphology.
The average SFH of red \ets with $SSFR<10^{-12}\,yr^{-1}$
  is instead very different from that of BSF \ets, and of all other subsamples: these galaxies were forming stars at a high rate at early epochs, then their SFR declined steeply with time. 
This effectively rules out the RP \ets as candidate progenitors for the BSF \ets. 

Moving the attention to RP \lts,  their SFH strongly depends on stellar mass (being steeper at higher masses) and does not resemble that of blue \lts, showing a much stronger variation with time. In the last $\sim$6 Gyr their average SFR is lower than that of blue \lts of similar mass.  
Most probably, the RP \lt population comprises all \lt galaxies that have stopped forming stars at any epoch and that have retained their morphology. Indeed, inspecting the SFH of individual galaxies, we find a number of RP \lts that have stopped forming stars at intermediate or even high redshifts, $\sim 30$\% at $z \geq 0.3$. Therefore, it is quite logical to expect an average star formation decline as the one observed: those \lts that stopped forming stars at an early epoch only contribute in this plot in the oldest age bins, making the decline slope steepen. In this sense, it is not logical to search for an evolutionary direct connection between all of today's RP \lts and today's blue \lts: only the subset of latest
arrivals of the former have recently evolved from the latter. The differences in the right panel of Fig.\ref{sfh} can be explained in such a scenario.

\section{Color evolution for different quenching histories}
\begin{figure*}
\centering
\includegraphics[scale=0.35]{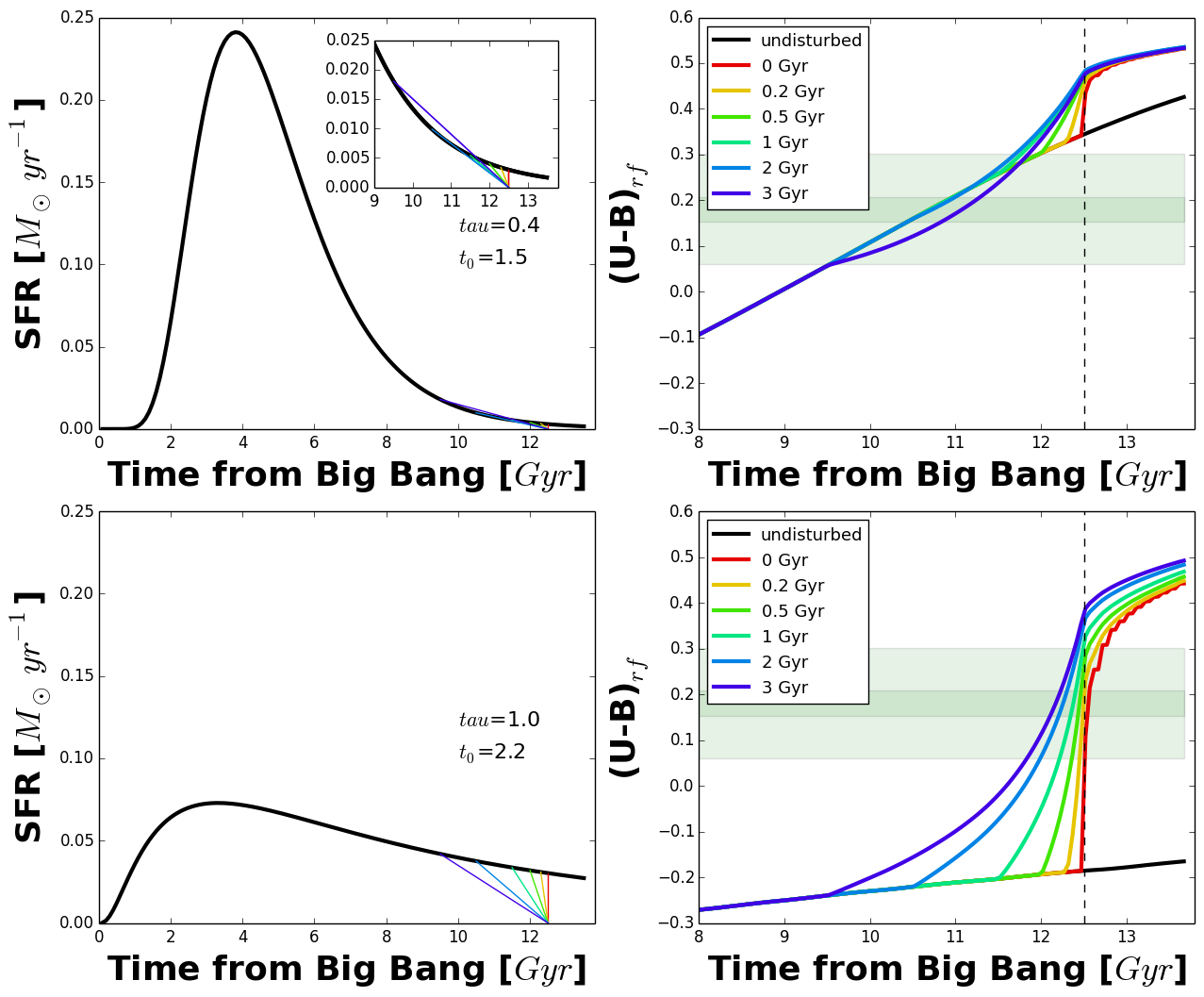}
\caption{Color of galaxies with different histories and different quenching timescales according to the spectrophotometric model of \citet{fritz07, fritz11}. Left panels: SFRs as a function of time with different $\tau$ and $t_0$, as written in the labels. The linear decline with different timescales is also shown with different colors. In the upper panel, a zoom of the decline is also shown.  Right panels: (U-B)$_{rf}$ color as a function of time for the SFRs shown in the left panels. The different declines are also shown. Green shaded areas show our definition of green galaxies (see \S\ref{populations}) for galaxies with \M=10.25 and \M=11.5.  \label{sfh_mod}}
\end{figure*}
To investigate what type of star formation history can produce a green galaxy spectrum, and for how long, we employ our spectrophotometric model described in \S\ref{data} \citep{fritz07, fritz11}   to compute the color of galaxies with different histories and different quenching timescales.
We use the SFH lognormal analytic form of \cite{gladders13}:
\begin{eqnarray}
SFR(t,t_0,\tau) =\frac{1}{t\sqrt{2\pi\tau^2}}\exp{-\frac{[\ln({t}-t_0)]^2 }{2\tau^2}}
\end{eqnarray}
where $t$ is the elapsed time since the Big Bang, $t_0$ is the logarithmic delay time, and $\tau$ sets the rise and decay timescale. This form
 has been shown to be a good representation of galaxy SFHs.

We consider galaxies with various lognormal parameters ($t_0$ and $\tau$) and compute their $(U-B)$ rest-frame color assuming: a) no quenching at any time; b) a sudden quenching, i.e. an abrupt interruption of the star formation activity on a timescale $t_{quench}=0$; c) a linear decline of the SFR from the undisturbed level to zero with short to long timescales: $t_{quench} = 0.2,0.5,1,2,3$ Gyr.
By construction, all the histories have a SFR equal to 0 1.1 Gyr ago. 

Figure \ref{sfh_mod} illustrates two emblematic cases of our modeling: one with a short $\tau$ (usually considered typical of early spirals) and one with a long $\tau$ (typical of late spirals). In the first case, galaxies can assume green colors for $\sim 2.5$ Gyr even without being subject to a quenching event, while in the second case intermediate colors are probably a sign of quenching. A short quenching time ($t_{quench}=0-0.5$ Gyr) produces a very short green phase ($\sim$0.15 Gyr), that might be hardly observable, while galaxies whose SFR declines with $t_{quench}=1-3$ Gyr go through a green phase that lasts for 0.25-0.5 Gyr. 

Galaxies with a very short $\tau$ ($\sim 0.3$, typical of \ets) do not go through a green phase at least  in the last 6 Gyr (plot not shown).

To conclude, our modeling shows that green colors are due to SFHs declining with long timescales. These could be due either to ``undisturbed'' lognormal histories with short $\tau$s, or to long $\tau$ galaxies
that are quenched on the timescales of the order of 1 Gyr or more. Green colors are therefore not necessarily indicative of ``quenching'' processes.

\section{Discussion}\label{disc}
In this paper we have investigated 
how galaxies of different stellar mass transform from one type to another in a variety of environments (groups, binary systems, single galaxies), in the local universe. We found that a non negligible fraction of objects is likely in a transitional phase, that is they are experiencing or have recently experienced a transformation from being star-forming to passive, or viceversa. We have seen that there is not a one-to-one correspondence between color and morphology, and confirmed that in many cases these two quantities must change on different timescales and might have a different dependence on stellar mass,  environment and galaxy class. 
We have then investigated the properties of the galaxies in transition, and compared them to those of their ``normal'' counterparts, to understand whether also their structural properties are different, or whether variations are limited to the efficiency of the SFR, which is mainly reflected in the galaxy color. 
Finally, we inspected  SFHs, to gain insights of the rate at which all galaxies have been forming stars in the past. 

\subsection{A possible evolutionary scenario}
\begin{figure}
\centering
\includegraphics[scale=0.37]{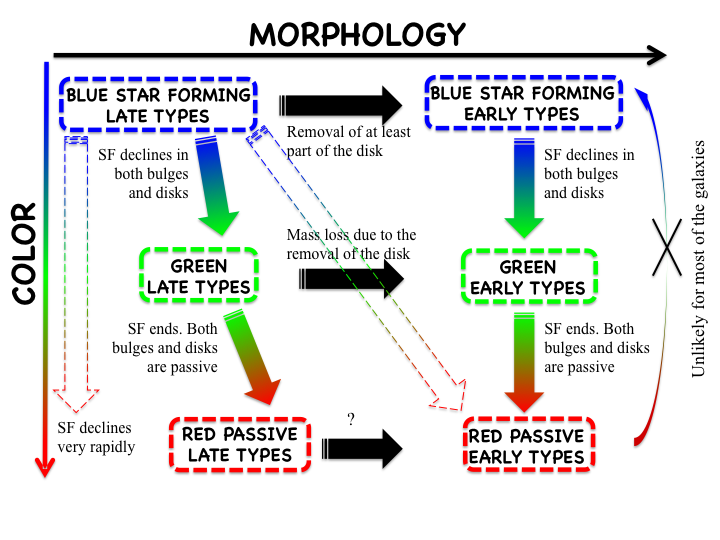}
\caption{Illustration of our main results and interpretation. \label{cartoon}}
\end{figure}

We are now in the position of proposing a possible evolutionary scenario for galaxies, as summarized in Figure \ref{cartoon}. 
\\
\\
{\it From blue \lts...}

We start considering blue (star-forming) \lt galaxies. At a certain epoch, because of secular and/or environmental processes, their gas supply is affected and their SFR starts to decline, either on a short or a long timescale. 
Therefore, a change in color must occur, either accompanied by a change in morphology or not. 
Various scenarios are possible at this stage.
\\
\\
{\it ...to green \lts}

If the SFR declines, blue \lts can turn into green \lts. Indeed, these two families share a similar SFH for most of the time, with green \lts showing a reduction of the SFR only in the last few Gyr (Fig.\ref{sfh}). 
Therefore, green \lts are most probably the result of blue \lts that have their SFR
suppressed, but not yet extinguished, both in the bulge and in the disk.
The analysis of their structural parameters is consistent with this picture, if a small structural variation takes place, 
in the sense of a small increase of the bulge relevance from blue to green. 
According to our modeling,  galaxies can assume green colors for $\sim$2.5 Gyr if their SFR is characterized by short $\tau$ even without being subject to any quenching event. Alternatively, galaxies whose SFR is characterized by long $\tau$  can remain green for up to 1 Gyr as a consequence of a quenching event.
\\
\\
{\it ... or to BSF \ets}

If the morphology changes before star formation is shut off, blue \lts
can turn into BSF \ets. 
The analysis of the SFHs, 
and their mass distributions support this hypothesis. Structural parameters of blue \lts and BSF \ets show significant differences instead, as expected if a morphological
transformation has occurred.  At any given mass, the disk in BSF \ets is not only smaller 
and fainter (different B/T ratio distribution) but also less massive,
suggesting that the morphological transformation happened with a removal of at least part of the stellar disk.  
\\
\\
{\it From green \lts...}

Subsequently,  green \lts  can further undergo to a change in color, and turn into RP \lts, or change their morphology and become green \ets. 
\\
\\
{\it ... to RP \lts}

If no major morphological transformation occurs,\footnote{A minor morphological transformation from late spirals to slightly earlier spirals is favored by the morphological distributions of green \lts vs. blue \lts.} green \lts most likely become RP \lts. Compared to their first progenitors (blue \lts) and their possible recent progenitors (green \lts), RP \lts are systematically more massive, and at a given stellar mass have a smaller disk.
These results, together with the analysis of SFHs,
suggest that RP \lts today can not derive only from the current blue and green \lts via the quenching and fading of a disk. They are probably an heterogeneous population which comprises all late type galaxies that have stopped forming stars at any epoch and that retained their morphology. 
The non negligible fraction of post-starburst galaxies  ($\sim15\%$) in RP \lts suggests that some of these objects became red after a very short (therefore virtually unobservable) green phase. 
\\
\\
{\it ... or to green \ets}

If green \lts are subject  to a morphological transformation, they might become green \ets. The analysis of the SFH supports such possibility, since no differences in SFH have been detected between the two populations (plot not shown). Green \ets are characterized by a steeper mass distribution than the green \lts, which extends toward slightly higher masses.  
The analysis of the structural parameters suggests that this mass loss is mainly related to the progressive disappearance of the disk, which at any given stellar mass is smaller and less massive.
The properties of the bulges of green \lts and green \ets are more similar, as a consequence the relative proportion of bulges and disk changes, as reflected in the distribution of  B/T ratios. 
\\
\\
{\it From BSF \ets  to green \ets}

Green \ets can also derive from BSF \ets which suffer a reduction of their SFR and a consequent change in color. 
Indeed, green and BSF \ets present very similar mass and structural parameters (\s indexes, B/T ratios, 
size of bulges and disks) distributions, 
supporting this scenario. 
\\
\\
{\it From green \ets and RP \lts to red \ets}

Finally, both green \ets and RP \lts can eventually turn into red \ets. 
However, given the fact that the RP \et population contains galaxies that stopped forming stars at any earlier epoch, which were characterized by different structural properties, comparisons between such populations are very hard and it is difficult to state the frequency of such transformations.  

In principle, red \ets might suffer a rejuvenation process that makes them change colors.
Indeed, some structural parameters of red and BSF \ets are similar, consistently with the hypothesis a common origin. However, this scenario seems to be ruled out by the characterization of the SFHs which are clearly different for the two populations. 
\\

Adopting a different galaxy selection, \cite{tojeiro13} found that red late-type spirals are recent descendants of blue late-type spirals, sharing similar SFHs at early times and showing a reduction of the star formation only in the last 500 Myr. On the basis of the SFH and dust content, they claimed red early-type spirals are more likely to evolve directly into red ellipticals than red late-type spirals. They also found that blue ellipticals show similar SFHs as blue spirals, except for a reduction of the efficiency of the star formation in the last 100 Myr. Blue ellipticals have different dust content than all spiral galaxies, ruling out the scenario according to which most of them  derive from  blue spirals.

\subsection{The physical processes responsible for galaxy transformations}
Different mechanisms have to be at work to produce the observed transformations.

Color transformations are due to a reduction and suppression of the SFR both in bulges and disks.
They can occur as the result of an external process or simply because galaxies use up all of their gas and shut down their star formation while they retain their structure. 

In the case of \lts, by taking a realistic accretion histories from cosmological simulations, \cite{forbes12} have shown that a certain fraction of disks in the course of their lifetimes is expected to experience a period of low accretion during which they will exhaust their gas supply and become redder, only to return to the blue cloud with the resumption of higher accretion rates. 

Ram pressure stripping of cold gas \citep{gunn72, feldmann11}, and strangulation of the galactic system by removal of hot and warm gas necessary to fuel star formation \citep{larson80, balogh00, font08} might be responsible for such transformations, since they are expected  not to affect  galaxy morphology, at least not directly. 
However, we do not detect any environmental dependence (see \S 7.3), suggesting these processes are not very efficient. In addition,  while ram pressure is observed to act on galaxies in high-mass clusters, it has not been observed in lower mass groups, where lower halo gas densities and satellite velocities likely lower its efficiency.
Moreover, haloes of $M_\ast \sim  10^{12} M_\odot$ are not expected to have virial shock fronts which support hot, virialized gas within the halo \citep{dekel06}, so in this mass regime it is not clear that either strangulation or ram pressure can be efficient. 
This may suggest tidal stripping  (e.g., \citealt{park07} and references therein), or  harassment and/or mergers  induce rapid cold gas consumption that quenches star formation. 
However, all these processes are also responsible for a morphological variation, since the distribution of the light and gas are also affected. 
Since the external parts of the galaxies first feel these processes, it is  plausible that they first affect disks, producing their observed fading. 
Therefore, a morphological change in the direction of an increase of the  B/T ratio might actually be an aftermath of quenching, simply due to the fading and the reduction of a star-forming disk once star-formation is reduced or ceased.

\subsection{Lack of environmental effects}

Our analysis has revealed very little evidence for environmental effects on galaxy transformations. 

The relative proportion of green and blue galaxies has been found to be  almost constant with environment (groups, binaries, satellites, and as a function of group-centric radius) and also for MMGs and satellites, implying that changes in color occur on a time scale that does not depend on host halo mass,
halo-centric radius or on being a central or a satellite in a halo.
In alternative, if most green galaxies are not due to quenching, the constancy of the green to blue ratio would indicate that the processes giving origin to long and short $\tau$ SFHs do not change efficiency with environment.
The decline in star formation giving rise to green galaxies is therefore not due to environmental effects. On the other hand, 
the relative fraction of green and blue galaxies depends on mass, being a factor 2:1 at high masses, 1:1 at intermediate masses, and $\sim$1:3 at lower
masses, suggesting that the decline in star formation in green galaxies is related to galaxy mass.

The only detected environmental effects are that 1) in single-MMGs, at any given mass, there are proportionally more blue and green galaxies than in the other environments, and in general they are proportionally less massive and 2) in groups there might be a higher fast quenching 
efficiency which gives origin to post-starburst signatures.

Our findings might be quite surprising, given that a host halo's virial radius broadly corresponds to a physical transition from the low-density field environment to a high-density region where dark matter and gas are virialized, therefore environmental effects would be expected. 

On the other hand, it is important to keep in mind that  environmental dependences can extend to galaxies beyond the virial radius of a group/cluster (as largley discussed in  \citealt{wetzel14}).
Therefore the (lack of) trends are plausibly driven, at least in part, by those galaxies having passed within much smaller distances from a group/cluster, but at the time of the observations are found very far from it. Also a large fraction of central galaxies near massive host haloes are actually ejected satellites (e.g., \citealt{wetzel14}).
\cite{baloghN00} first noted from N-body simulations of clusters that particles that have passed within the virial radius can then orbit well outside of it. 
`ejected' (or `backsplash') satellites (see e.g, \citealt{Gill05, bahe13})  typically orbit back out to a maximum distance of $\sim$2.5 R$_{200}$ beyond a host halo after passing through it. In addition, almost half of all galaxies within this distance are composed of these ejected satellites, which are preferentially quiescent (e.g. \citealt{wang09, wetzel13}),  with higher fractions for less massive galaxies and around more massive host haloes. Thus, ejected satellites are potentially critical for understanding the properties of galaxies near groups/clusters and obtaining a complete picture of environmental dependence. 

\subsection{The importance of the galaxy structure}
Recently, the importance of galaxy structure for galaxy transformations has been widely discussed in the literature.

\cite{carollo14} report that the quenched satellites at low $z$ have larger B/T and smaller half-light radii than the star-forming satellites. 
They find that differences are mostly due to differences in the disks, which have lower luminosities in the quenched galaxies, but they can not be explained by uniformly fading the disks following quenching. Instead, either there must be a differential fading of the disks with galaxy radius or  disks were generally smaller in the past, both of which would be expected in an inside-out disk growth scenario. 

Other literature results focused mainly on the bulge, finding that its properties might play a role in quenching galaxies, rather than the disk. For example, in the local universe, \cite{abramson14} showed that the increase in bulge mass-fractions, which are portions of a galaxy not forming stars, is responsible for the existing anti-correlation between $SSFR$ and $M_\ast$.
At $z<0.2$, the passive fraction for central galaxies has been found to be closely correlated to the bulge mass \citep{bluck14} and to the B/T ratio \citep{omand14}. 
At 0.5$<z<$2.5 \cite{lang14} found an increased bulge prominence among quiescent galaxies, with an increase of the typical  B/T  among star- forming galaxies above $10^{11}M_\odot$.These findings have lead some authors to suggest that the physical mechanisms responsible for the quenching of star formation must be strongly coupled to the formation or accretion of the bulge.

Our analysis confirms that the relative importance of the bulge and the disk appears to be a key parameter in galaxy transformations.   
We found that a {\it morphological} transition (from BSF \lts to BSF \ets, from green \lts to green \ets) is mainly due to the fading and total or partial removal of the disk. 
A transition in star formation with no morphological change, instead (from blue \lts to green \lts) is accompanied by a small structural change: as we have seen in \S5.3, the average B/T ratio and mass ratio of bulge and disk in green \lts is only slightly higher than in blue \lts. In this case the differences in these ratios can be ascribed to bulge growth instead of disk fading, but the effect is overall small on the population.

\subsection{Quenching times}
According to our modeling, galaxies which sustain a green color for a non-negligible interval of time ($>0.5$ Gyr) can originate either from a long timescale ($>1$ Gyr) quenching of long $\tau$ galaxies, or from short $\tau$ "undisturbed" star formation histories typical of intermediate types. 
Galaxies with a short quenching (0-0.5 Gyr) timescale do exist, but they sustain green colors for a short period of time and they are observable as $k+a$ galaxies that quickly transit from being blue to being red. 

We have shown that the occurrence of green galaxies (compared to that of blue galaxies) does not depend on environment while that of $k+a$ does. This suggests that the {\it environmental quenching timescale} is short, while other galaxies go from being star-forming to being passive on a long timescale independently of environment.

We have seen that green \lts are twice as numerous as $k+a$'s.
Starting from the logical assumption that both green and $k+a$ galaxies have a common progenitor among star-forming galaxies, and if we assume that the green phase lasts for about twice the time (of the order of 2 Gyr) of the $k+a$ visibility ($\sim$1 Gyr), we conclude that the short timescale and the long timescale
SF ``quenching'' channels contribute about equally to the growth of the passive population.

Several authors have tried
to estimate the "quenching timescale" 
using different approaches. A direct comparison is impossible because it strongly depends on how a
"quenching or quenched" galaxy is defined. However, we report here some of the latest results as comparison reference.

\cite{wetzel13} found that satellite quenching is the dominant process for building up all quiescent galaxies at $M_\ast<10^{10}M_\odot$. They proposed a ``delayed-then-rapid'' quenching scenario: satellite SFRs is unaffected for 2-4 Gyr after infall, after which star formation quenches rapidly, with an e-folding time of $<$0.8 Gyr (see also \citealt{mcgee09, mcgee11, delucia12}). These quenching time-scales are shorter for more massive satellites but do not depend on host halo mass.

Investigating objects in transition defined using a color-color diagram, \cite{mok13} proposed a much shorter quenching time scale with an e-folding time of $<$0.5 Gyr.
\cite{schawinski14},  inspecting morphologies in addition to colors, concluding that only a small population of blue \ets move rapidly across the green valley after the morphologies are transformed from disk to spheroid and star formation is quenched rapidly ($\tau <0.25$ Gyr). In contrast, the majority of BSF galaxies have significant disks, and they retain their late-type morphologies as their star formation rates decline very slowly ($\tau >1$ Gyr). 

Only \cite{wheeler14}, studying a sample of dwarf galaxies, found much longer timescales  ($>$ 9.5 Gyr, a ``slow starvation'' scenario), concluding that the environmental processes triggering quenching must be highly inefficient.

\section{Summary and Conclusions}

Investigating a mass complete sample of galaxies drawn from the  PM2GC \citep{rosa}, this paper focused mainly on two points: 1) characterizing the color (red, green, blue) and morphological (ellipticals, S0s, \lts) transformation of galaxies as a function of the stellar mass and the environment and 2) studying the properties of the objects that are most likely in a transitional phase (green, RP \lts, BSF \ets), with the aim of understanding the evolutionary links between the different sub-populations.

Our analysis showed that the relative importance of the bulge and the disk seems to play an important role in galaxy transformations.   
We found that the fading and total or partial removal of the disk produces a morphological transition, so that BSF \lts and green \lts become BSF \ets and green \ets, respectively.
On the other hand, a transition can occur even without a noticeable structural and morphological change: SFR can declines both in bulges and disks, producing a variation in color. In this way blue galaxies turn into green (when they are still forming stars, but a reduced rate) and red galaxies.
Therefore, RP \lt galaxies descend from blue \lts that have stopped forming stars at any epoch (retaining their morphology),  
going through either a short or a long green phase. 

Our spectrophotometric model allowed us to better characterize the occurrence and duration of the green phase.
In some cases galaxies can turn from blue to red quite quickly, going through a very short green phase (<0.1 Gyr) hardly observable when considering only colors, but recognizable by their spectral $k+a$ features.
In other cases green colors are  not indicative of ``quenching'' processes. They are due to star formation histories declining with long timescales. These could be due either to ``undisturbed'' lognormal histories with $\tau$ typical of early-spirals or to long $\tau$ typical of late spirals that are quenched on the timescales of the order of at least 1 Gyr. 

\section*{Acknowledgements}
We thank the anonymous referee whose comments helped us to improve the readability of the paper. 
This work was supported by the World Premier International Research Center Initiative (WPI), MEXT, Japan.  It was also supported  by the
Kakenhi Grant-in-Aid for Young Scientists (B)(26870140) from the Japan Society for the Promotion of Science (JSPS).
The Millennium Galaxy Catalogue consists of imaging data from the
Isaac Newton Telescope and spectroscopic data from the Anglo
Australian Telescope, the ANU 2.3m, the ESO New Technology Telescope,
the Telescopio Nazionale Galileo and the Gemini North Telescope. The
survey has been supported through grants from the Particle Physics and
Astronomy Research Council (UK) and the Australian Research Council
(AUS). 

\bibliographystyle{apj}
\bibliography{biblio_groups}

\appendix

\section{Trends with \s indexes}\label{sersic}
\begin{figure*}
\centering
\includegraphics[scale=0.3]{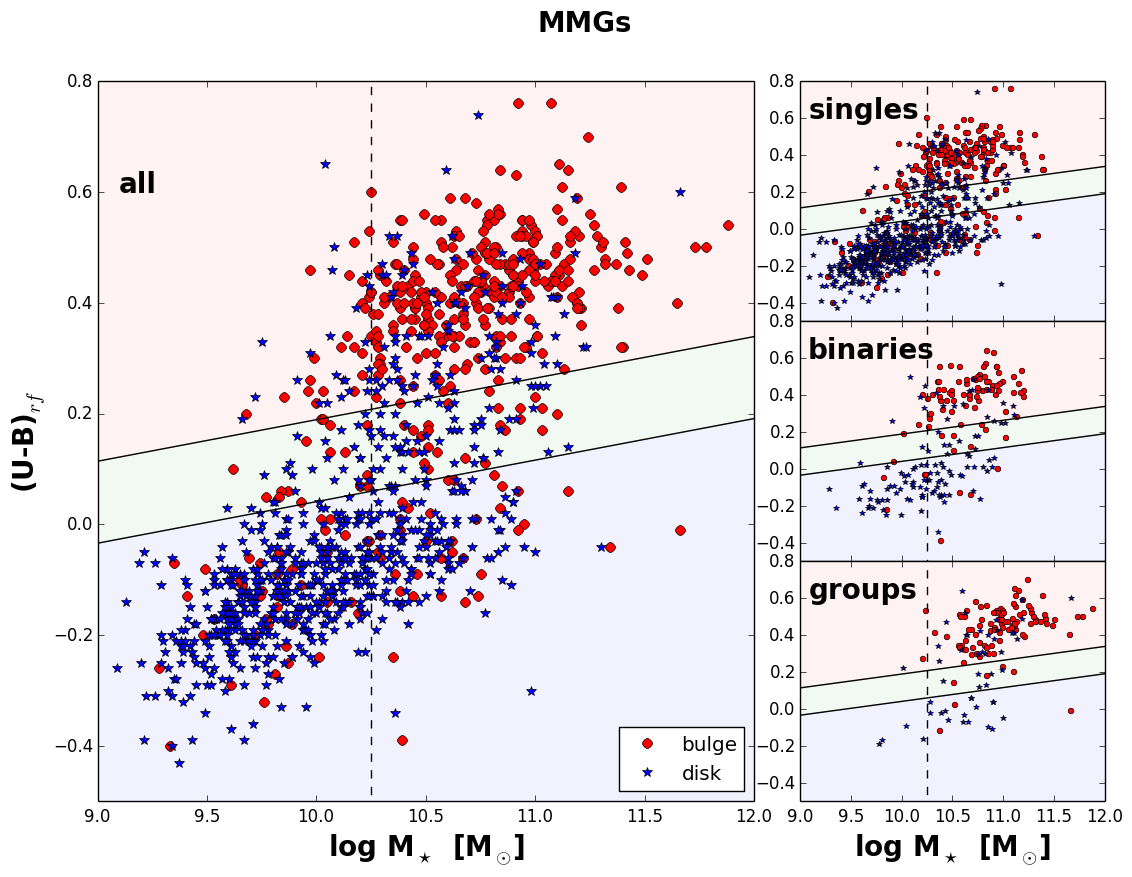}
\includegraphics[scale=0.3]{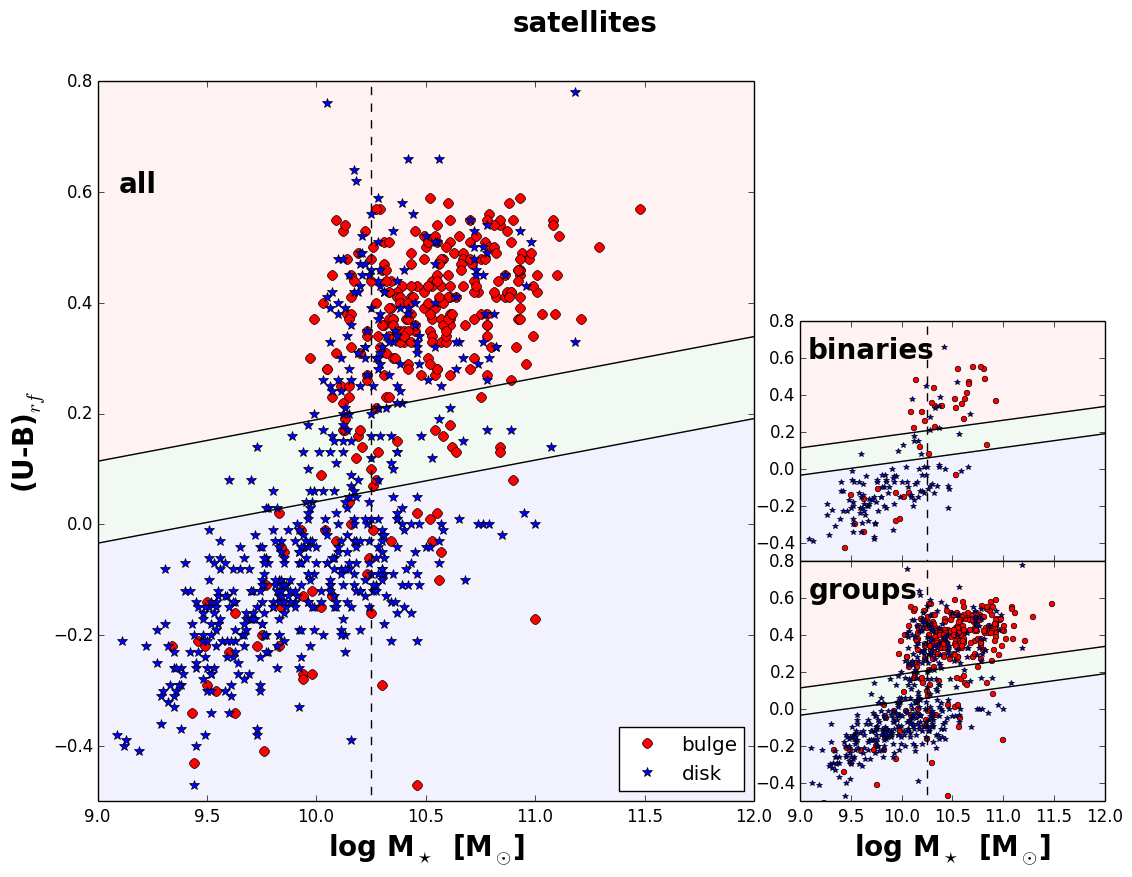}
\includegraphics[scale=0.3]{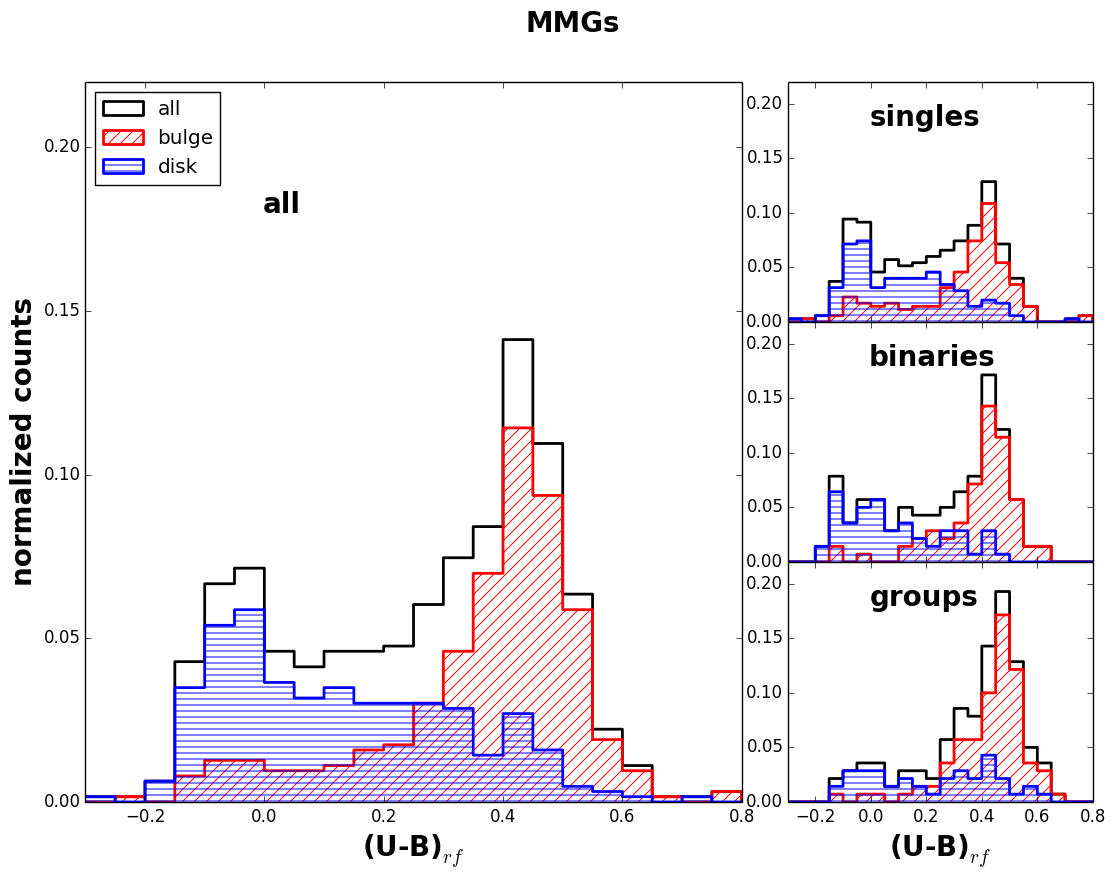}
\includegraphics[scale=0.3]{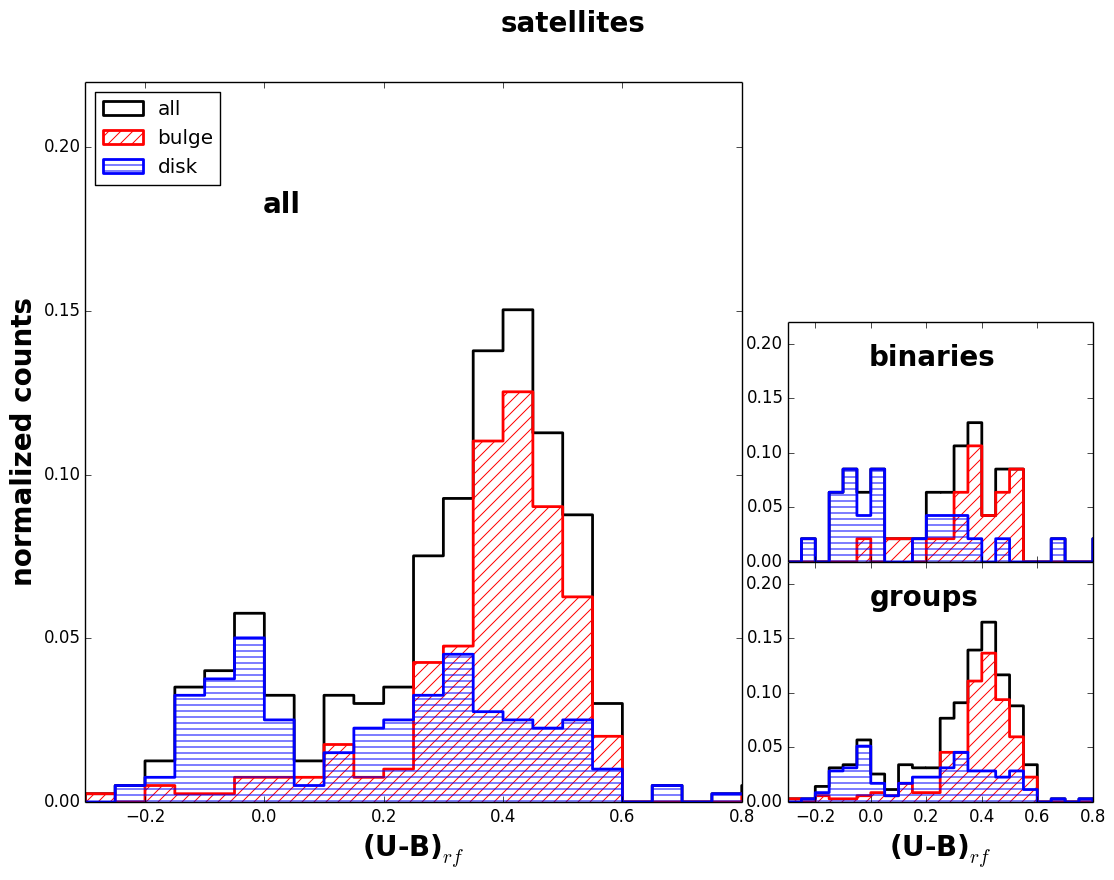}
\caption{Rest frame $(U-B)_{rf}$- mass relation (upper panels) and rest frame $(U-B)_{rf}$ color distribution (bottom panels) for galaxies of different types (left panels: MMGs , right panels: satellites) and \s indexes in the different environments (shown in the smaller windows). Red lines and circles: bulges, blue lines and stars: disks. In the upper panels, the black dashed vertical line represents the mass completeness limit, the black solid line shows the separation between red, green and blue galaxies \label{fig_col_ser}}
\end{figure*}

In this paper we have investigated the relationship between color and morphology. In this Appendix we show that adopting the \s index instead of morphology to distinguish between the sub-populations overall gives similar results, even though the analysis of the morphologies is more detailed. 

We separate the sample into bulge-dominated galaxies ($n>2.5$ -hereafter simply ``bulges'') and disk-dominated galaxies ($n<2.5$ - hereafter simply ``disks'').  
\begin{table}
\caption{Percentage of galaxies of different types above the stellar mass completeness limit in different environments. \label{tab_frac_ser}}
\centering
\setlength{\tabcolsep}{3pt}
\begin{tabular}{ll|cc cc cc cc}
\hline
\hline
&& \multicolumn{8}{c}{{\bf MOST MASSIVE GALAXIES}} \\
\hline
 && \multicolumn{2}{c}{{\bf ALL GALAXIES}}  & \multicolumn{2}{c}{{\bf SINGLE GALAXIES}} & \multicolumn{2}{c}{{\bf BINARY SYSTEMS}} & \multicolumn{2}{c}{{\bf GROUPS}}  \\
 & &  $bulge$ & $disk$ &  $bulge$ & $disk$ &  $bulge$ & $disk$ &  $bulge$ & $disk$ \\ 
\hline
\parbox[t]{2mm}{\multirow{5}{*}{\rotatebox[origin=c]{90}{{\bf COLOR}}}} & & \\
& $all$ 	&  54$\pm$3 & 46$\pm$3 	& 49$\pm$4 & 51$\pm$4 & 56$\pm$6  & 44$\pm$6 & 67$\pm$6 & 33$\pm$6\\ 
 & $red$ 	&  45$\pm$3 & 11$\pm$2 	& 37$\pm$4 & 15$\pm$3 & 48$\pm$6  & 10$\pm$4 & 62$\pm$6 & 17$\pm$5\\ 
 & $green$ 	&  3$\pm$1 	& 9$\pm$2	& 3$\pm$1  & 12$\pm$2 & 5$\pm$3   & 8$\pm$4 & 3$\pm$2 & 4$\pm$3\\
 & $blue$ 	&  5$\pm$1 	& 21$\pm$2	& 8$\pm$2  & 23$\pm$3 & 2$\pm$2   & 26$\pm$5 & 2$\pm$2 & 11$\pm$4\\
\hline
\hline
&& \multicolumn{8}{c}{{\bf SATELLITES}} \\
\hline
 && \multicolumn{2}{c}{{\bf ALL GALAXIES}}  & \multicolumn{2}{c}{{\bf SINGLE GALAXIES}} & \multicolumn{2}{c}{{\bf BINARY SYSTEMS}} & \multicolumn{2}{c}{{\bf GROUPS}}  \\
 & &  $bulge$ & $disk$ &  $bulge$ & $disk$ &  $bulge$ & $disk$ &  $bulge$ & $disk$ \\ 
\hline
\parbox[t]{2mm}{\multirow{5}{*}{\rotatebox[origin=c]{90}{{\bf COLOR}}}}& & \\
& $all$	 	& 57$\pm$3 & 43$\pm$3 &--&--& 47$\pm$11 & 53$\pm$11& 59$\pm$4 & 42$\pm$4\\ 
 & $red$ 	& 51$\pm$4 & 23$\pm$3 &--&--& 40$\pm$11  & 19$\pm$9 & 52$\pm$4 & 22$\pm$3\\ 
 & $green$	& 3$\pm$1  & 5$\pm$2 &--&--&  4$\pm$5   & 4$\pm$5  &  3$\pm$1 & 5$\pm$2\\ 
 & $blue$	& 3$\pm$1  & 16$\pm$2 &--&--&  2$\pm$4  & 30$\pm$10 &  3$\pm$1 & 14$\pm$3\\ 
\hline
\hline
\end{tabular}
\end{table} 
Figure \ref{fig_col_ser} shows the same as Fig.\ref{fig_col}. Bulges are most likely red and more massive, while disks tend to be bluer. As shown in Tab.\ref{tab_frac_ser}, bulge and disk fractions depend on the environment:  for both MMGs and satellites, bulges are more common in groups than in single systems. However, in a given environment,  fractions differ for satellites and MMGs only in groups, with MMGs hosting a higher fraction ($\sim$70\%) of bulges than satellites ($\sim$60\%). 
Both for MMGs and satellites and in all environments, the distribution of disks resembles the one presented for \lts and that of bulges that of the sum of ellipticals+S0s.

\begin{figure*}
\centering
\includegraphics[scale=0.35]{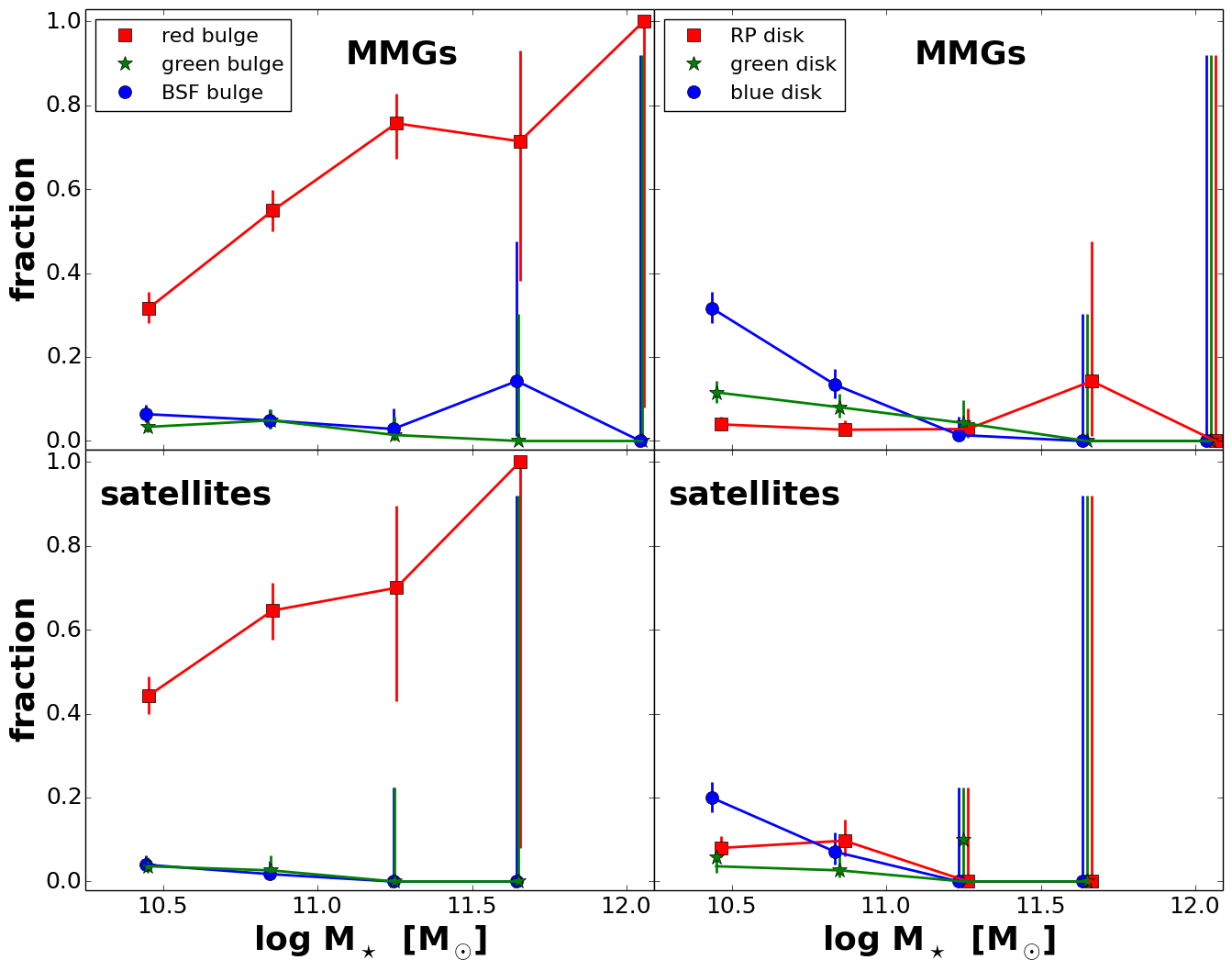}
\caption{Fraction of galaxies as a function of stellar mass for MMGs (upper panels) and satellites (bottom panels), when a color+SSFR+\s index cut is adopted. Left panels: \ets, as indicated in the labels. Right panels: \lts, as indicated in the labels. 
Errors are defined as binomial errors \citep{gehrels86}.
\label{col_frac_mass_dist_ser}}
\end{figure*}

Figure \ref{col_frac_mass_dist_ser} shows the same as Fig.\ref{col_frac_mass_dist}. In both MMGs and satellites,  blue and green disks show similar trends to blue and green \lts, that is they are more frequent at low masses than at higher. In MMGs, together, they dominate the total population at low masses, while in satellites red bulges always dominate, with their importance increasing with increasing mass. In general, red bulges resemble red \ets. 
We note that in satellites the contribution of red bulges increases from 40\% to 100\% without showing any dip. This might suggest that the weird trend seen in red early and \lts is due to galaxies that are morphologically classified as late, but have a non negligible bulge. 
In both classes, trends for BSF bulges, RP disks and green bulges are almost flat. All these galaxies are absent only in the highest mass bin. 
Each sub-population represents $\leq$15\% of the total population.  

\begin{figure*}
\centering
\includegraphics[scale=0.26]{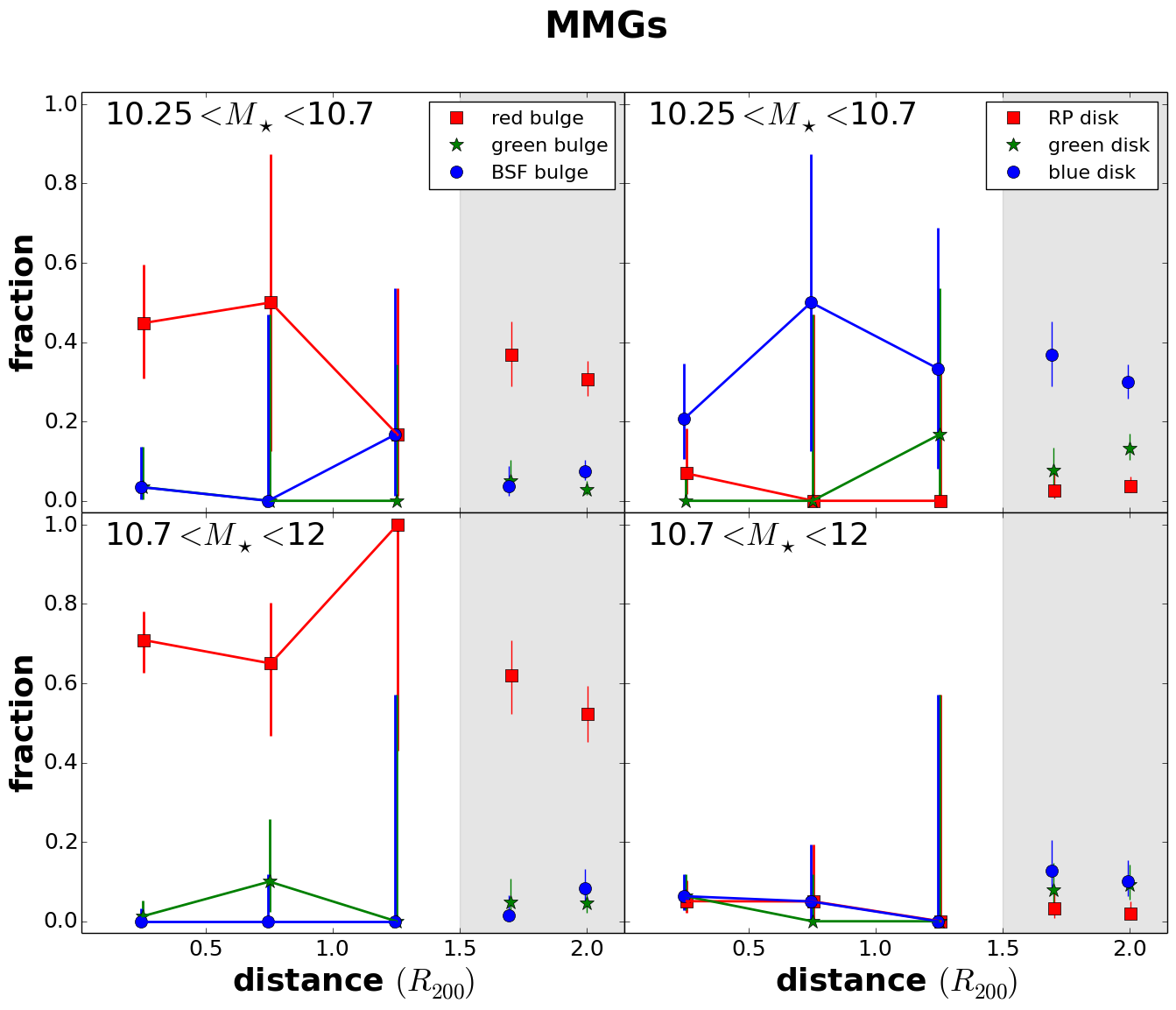}
\includegraphics[scale=0.26]{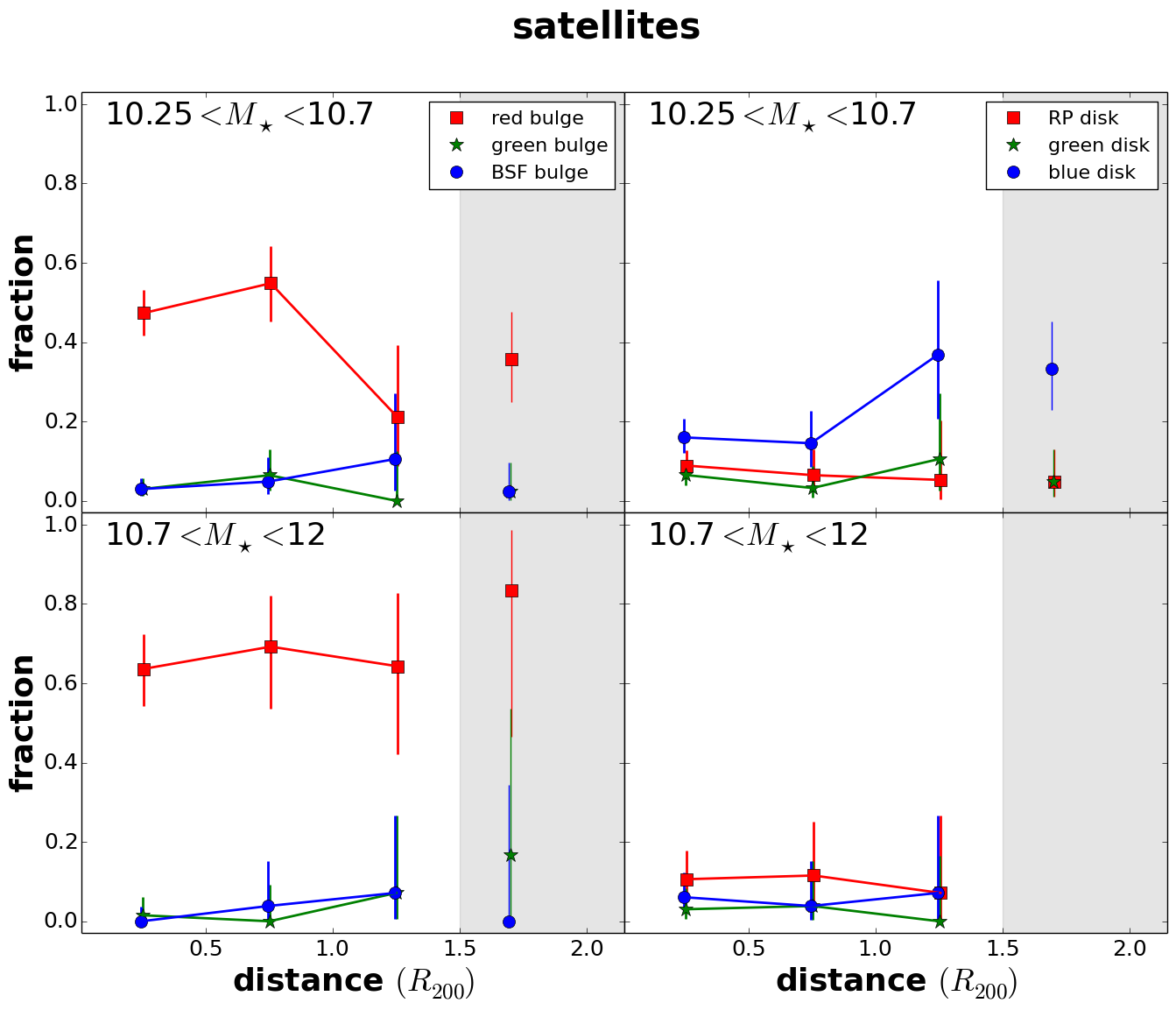}\caption{Fraction of galaxies as a function of group-centric distances in bins of stellar mass, when a color+SSFR+\s index cut is adopted. Left panel: MMGs. Right panel: satellites. Left panels: \M$<$10.7, right panel: \M$>$10.7. Colors and symbols are as in Fig.\ref{col_frac_mass_dist_ser}. Grey area represent galaxies that are not in groups, i.e. binary systems (placed at $r/R_{200}$=1.7) and single galaxies  (placed at $r/R_{200}$=2). Errors are defined as binomial errors \citep{gehrels86}. \label{col_frac_dist_sersic}}
\end{figure*}

Focusing on groups, Figure \ref{col_frac_dist_sersic} shows the same as Figure \ref{col_frac_dist}. Both at low and high masses, within the errors, trends for satellites and MMGs are compatible. At low masses, the incidence of red bulges (which dominate close to the group centers) seems to decrease at large distances. In contrast,  the fraction  of blue and green disks and BSF disks slightly increases. Green bulges are almost absent at all distances. 
At higher masses, both in MMGs and satellites, red bulges dominate at all distances  ($>$60\%), and their fraction increases with distance. In contrast, the other three classes of objects show slightly declining  trends. 
In the non group environments, in both mass bins fractions are similar to those of group outskirts, given the large uncertainties.

\section{Ellitpicals and S0s}\label{S0ell}
In the main text we have presented results for ellipticals and S0s together, nonetheless it is known that they are characterized by different properties. Table \ref{tab_trans_ellS0} presents the characteristic numbers for the two populations.

Red ellipticals and S0s differ especially in the B/T ratio, with the former being more bulge dominated than the latter. Similar discrepancies are found also for the objects in transitions. 
In addition, we note that BSF ellipticals are more star-forming than S0s, showing on average higher values of SFR and SSFR.

\begin{table}
\caption{Characteristic numbers of ellipticals and S0s, compared to normal galaxies of the same morphology. 
\label{tab_trans_ellS0}}
\centering
\begin{tabular}{lcccccc}
\hline
\hline
Quantity &\multicolumn{3}{c}{{\bf ellitpicals}} &\multicolumn{3}{c}{{\bf S0s}} \\
		 & red 	& green & BSF & red & green & BSF 	 \\
\hline
number & 295 & 15& 18 & 281 & 36 & 32\\
$\langle$B$_{Vega}$ $\rangle$ &-19.97$\pm$0.05&-19.8$\pm$0.2 &-20.4$\pm$0.1 &-19.87$\pm$0.05& -19.8$\pm$0.1 &-20.28$\pm$0.09 \\
$\langle$\M$\rangle$  &10.66$\pm$0.05&10.4$\pm$0.1&10.40$\pm$0.06&10.62$\pm$0.05& 10.48$\pm$0.06 &10.39$\pm$0.07\\
$\langle$SFR$\rangle$($M_\odot yr^{-1}$) &0$\pm$0&1$\pm$1&4.7$\pm$0.7&0$\pm$0& 1.8$\pm$0.5 & 3$\pm$1\\
$\langle$SSFR$\rangle$$^{\alpha}$ ($yr^{-1}$)&(8$\pm$5)10$^{-12}$&(3$\pm$3)10$^{-11}$&(1.5$\pm$0.2)10$^{-10}$&(9$\pm$3)10$^{-12}$& (4$\pm$1)10$^{-11}$ & (7$\pm$3)10$^{-11}$\\
$\langle$n$\rangle$  &3.4$\pm$0.1&2.8$\pm$0.4&2.3$\pm$0.5&3.3$\pm$0.1& 2.7$\pm$0.4 & 1.8$\pm$0.4\\
$\langle$B/T$\rangle$ &0.61$\pm$0.02&0.58$\pm$0.08&0.7$\pm$0.1&0.51$\pm$0.02& 0.52$\pm$0.06 & 0.25$\pm$0.08\\
$\langle R_e \, (bulge)\rangle$(kpc) &1.13$\pm$0.07&0.9$\pm$0.2&1.2$\pm$0.3&1.09$\pm$0.08& 1.2$\pm$0.3 & 1.2$\pm$0.2\\
$\langle R_e \, (disk)\rangle$ (kpc) &2.4$\pm$0.1&2.2$\pm$0.4&1.5$\pm$0.2&2.1$\pm$0.1& 2.0$\pm$0.2 & 2.0$\pm$0.2\\
$\langle$(u-r) bulge$\rangle$$^\beta$  &2.66$\pm$0.04&2.2$\pm$0.1&1.95$\pm$0.09&2.65$\pm$0.01& 2.3$\pm$0.1 & 1.91$\pm$0.06\\
$\langle$(u-r) disk$\rangle$$^\beta$ &2.52$\pm$0.05&2.2$\pm$0.3&2.0$\pm$0.3&2.53$\pm$0.06& 2.3$\pm$0.1 & 1.9$\pm$0.1\\
\hline
\multicolumn{7}{l}{\footnotesize{$^{\alpha}$Values computed only with galaxies with SSFR$\neq$0}}\\
\multicolumn{7}{l}{\footnotesize{$^{\beta}$Colors are in the AB system}}\\
\end{tabular}
\end{table}

\label{lastpage}
\end{document}